\documentclass[iop]{emulateapj}
\usepackage{multirow,rotating,lscape}

\newcommand{\kms}{\mbox{km~s$^{-1}$}}

\shorttitle{Millimeter observations of NGC~1068} 

\shortauthors{Krips et al.}

\begin{document}


\title{SMA/PdBI multiple line observations of the nearby Seyfert~2
  galaxy NGC~1068:\\ Shock related gas kinematics and heating in the
  central 100~pc?$^\star$}\thanks{$^\star$ Based on observations
  carried out with the IRAM Plateau de Bure Interferometer. IRAM is
  supported by INSU/CNRS (France), MPG (Germany) and IGN (Spain).}

\author{M. Krips\altaffilmark{1}, 
S. Mart\'{\i}n\altaffilmark{2},
A. Eckart\altaffilmark{3},
R. Neri\altaffilmark{1},
S. Garc\'{\i}a-Burillo\altaffilmark{4}, 
S. Matsushita\altaffilmark{5,6},
A. Peck\altaffilmark{7,6},
I. Stoklasov\'a\altaffilmark{8},
G. Petitpas\altaffilmark{9}, 
A. Usero\altaffilmark{4},
F. Combes\altaffilmark{10},
E. Schinnerer\altaffilmark{11},
L. Humphreys\altaffilmark{12},
A.J. Baker\altaffilmark{13}
}

\altaffiltext{1}{Institut de Radio Astronomie Millim\'etrique, Saint
Martin d'H\`eres, F-38406, France; email: [krips,neri]@iram.fr}

\altaffiltext{2}{European Southern Observatory, Alonso de C\'ordova
3107, Vitacura, Casilla 19001, Santiago 19 Chile; e-mail:
smartin@eso.org}

\altaffiltext{3}{Universit\"at zu K\"oln, I.Physikalisches Institut,
Z\"ulpicher Str. 77, 50937 K\"oln, Germany; email:
        eckart@ph1.uni-koeln.de}

\altaffiltext{4}{Observatorio Astron\'omico Nacional (OAN) -
Observatorio de Madrid, C/ Alfonso XII 3, 28014 Madrid, Spain;
email: [s.gburillo,a.usero]@oan.es}

\altaffiltext{5}{Institute of Astronomy and Astrophysics, Academia
Sinica, PO Box 23-141, Taipei 10617, Taiwan, R.O.C.; email:
satoki@asiaa.sinica.edu.tw}

\altaffiltext{6}{Joint ALMA Observatory, Alonso de C\'ordova 3107,
  Santiago, Chile; email: apeck@alma.cl}

\altaffiltext{7}{NRAO, 520 Edgemont Rd, Charlottesville, VA 22903}

\altaffiltext{8}{Astronomical Institute of the Academy of Sciences of
the Czech Republic, v.v.i., Bo\v{c}n\'{\i} II 1401, 14131 Prague, Czech
Republic}

\altaffiltext{9}{Harvard-Smithsonian Center for Astrophysics, SMA
project, 60 Garden Street MS 78, Cambridge, MA 02138; email:
gpetitpa@cfa.harvard.edu}

\altaffiltext{10}{Observatoire de Paris, LERMA, 61 Av. de
  l'Observatoire, 75014 Paris, France; email:
  francoise.combes@obspm.fr}

\altaffiltext{11}{Max-Planck-Institut f\"ur Astronomie, K\"onigstuhl
17, 69117 Heidelberg, Germany; email: schinner@mpia.de}

\altaffiltext{12}{ESO, Karl-Schwarzschild-Str. 2, D-85748 Garching,
Germany; email: ehumphre@eso.org}

\altaffiltext{13}{Department of Physics and Astronomy, Rutgers, the
State University of New Jersey, 136 Frelinghuysen Road, Piscataway, NJ
08854-8019, USA; email: ajbaker@physics.rutgers.edu}

\begin{abstract}
We present high angular resolution ($0\farcs5-2\farcs0$) observations
of the mm continuum and the $^{12}$CO(J=3--2), $^{13}$CO(J=3--2),
$^{13}$CO(J=2--1), C$^{18}$O(J=2--1), HCN(J=3--2), HCO$^+$(J=4--3) and
HCO$^+$(J=3--2) line emission in the circumnuclear disk
(r$\lesssim$100~pc) of the proto-typical Seyfert type-2 galaxy
NGC~1068, carried out with the Submillimeter Array. We further include
in our analysis new $^{13}$CO(J=1--0) and improved $^{12}$CO(J=2--1)
observations of NGC~1068 at high angular resolution
($1\farcs0-2\farcs0$) and sensitivity, conducted with the IRAM Plateau
de Bure Interferometer. Based on the complex dynamics of the molecular
gas emission indicating non-circular motions in the central
$\sim$100~pc, we propose a scenario in which part of the molecular gas
in the circumnuclear disk of NGC~1068 is radially blown outwards as a
result of shocks. This shock scenario is further supported by quite
warm (T$_{\rm kin}$$\geq$200~K) and dense
(n(H$_2$)$\simeq$10$^4$~cm$^{-3}$) gas constrained from the observed
molecular line ratios. The HCN abundance in the circumnuclear disk is
found to be [HCN]/[$^{12}$CO]$\approx$10$^{-3.5}$. This is slightly
higher than the abundances derived for galactic and extragalactic
starforming/starbursting regions. This results lends further support
to X-ray enhanced HCN formation in the circumnuclear disk of NGC~1068,
as suggested by earlier studies. The HCO$^+$ abundance
([HCO$^+$]/[$^{12}$CO]$\approx$10$^{-5}$) appears to be somewhat lower
than that of galactic and extragalactic starforming/starbursting
regions. When trying to fit the cm to mm continuum emission by
different thermal and non-thermal processes, it appears that
electron-scattered synchrotron emission yields the best results while
thermal free-free emission seems to over-predict the mm continuum
emission.

\end{abstract}


\section{Introduction}
Little is known about the effects of active processes in galaxies on
the chemical and kinematic properties of the surrounding molecular gas
and vice versa, whether the activity is in form of an active galactic
nucleus (AGN) or a starburst (SB) or both. Information on the
characteristics of the molecular gas in the vicinity of the activity
is essential to reveal the underlying physical processes because
molecular gas constitutes a large fraction of the fuel for the central
activity and thus helps to keep it alive over cosmologically relevant
time scales. Also, the feedback of activity onto the surrounding
molecular gas represents an important factor for the evolution of the
activity (with respect to outflows or shocks for instance).  The large
diversity and oftentimes also simultaneity of the physical processes
accompanying the different activity types certainly complicate any
interpretation of the interaction between the activity and the
molecular gas. These processes include large scale shocks, gas out-
and inflow, other dynamical perturbations, and strong radiation
fields, such as through UV- or X-ray radiation, cosmic rays, or
supernovae explosions \citep[e.g.,][]{mart11, garc10, saka10, papa10,
perez09, krips08, aalt08, mats07, garc07, mart06, user06, saka06,
mats05, fuen05, mei05, user04}. Thus, a thorough study of the
kinematics, excitation conditions {\it and} chemistry of the molecular
gas close to AGN and SBs is essential for understanding the nature and
evolution of these active environments.

High angular resolution observations of the $^{12}$CO emission, a
reliable tracer of the global molecular gas reservoir, are an
important step to study the dynamics in active galaxies. $^{12}$CO
alone, however, can certainly not describe the complexity of the
molecular gas close to the activity processes (especially with respect
to the chemistry and excitation conditions of the molecular
gas). Moreover, $^{12}$CO has been found to be an unreliable tracer of
dense molecular gas environments (at least its lower rotational
transitions), in which most AGN or star formation activity is supposed
to take place \citep[e.g.,][]{krip07,gao04}. Observations at high
angular resolution of different molecular tracers is thus a next
logical step. Given recent upgrades of current (sub)mm
interferometers, the detection and spatial resolution of weaker
molecular lines becomes a feasible task.

In this paper we present high angular resolution observations of the
(sub-)mm-continuum and $^{12}$CO, $^{13}$CO, C$^{18}$O, HCN and
HCO$^+$ line emission in the nearby Seyfert type-2 galaxy NGC~1068,
conducted with the Submillimeter Array (SMA) and the IRAM Plateau de
Bure Interferometer (PdBI).


\section{NGC~1068}
\label{1068}
The nearby Seyfert type-2 galaxy NGC~1068 (see Table~\ref{tab1} for a
general overview) has become not only the figurehead for the viewing
angle unification theory for Seyfert galaxies \citep[e.g.][]{krol87},
as its centre most impressively exhibits the characteristics for
harboring an obscured type-1 Seyfert AGN \citep[][]{anto85}. It also
advanced to a figurehead for the significantly different effects that
AGN can have on the excitation conditions and chemistry of the
surrounding molecular gas when compared to SB or quiescent galaxies
\citep[e.g.][]{krips08,user04,kohno01,ster94}. The prototypical nature
of NGC~1068 is certainly in part due to its relatively small distance
(12.6~Mpc) from us and strong continuum as well as line emission from
X-ray to radio frequencies, making it hence an ideal target to study
the accretion and feedback processes of its active nucleus at
unprecedented detail. As a (fortunate) consequence, a wealth of
information is already available for this source, the most revelant of
which will be summarised in this section.

The molecular gas in NGC~1068 is distributed in a starburst
ring/spiral of $\sim$3~kpc ($\sim$50$''$) in diameter, a stellar bar
of $\sim$2~kpc ($\sim$30$''$) in length and a cirumnuclear disk/ring
(CND) of $\sim$200~pc ($\sim$3$''$) in diameter \citep[e.g.,][and
references therein]{schin00}. At the very center of the CND,
pronounced H$_2$0 maser emission suggests a thin disk of a few parsec
in diameter \citep[e.g.,][]{gall01,gree96}. A pronounced jet and
counter-jet can be observed from cm to mm wavelengths
\citep[e.g.,][]{krip06,gall04} that extends from the maser disk out to
several kpc from the center. MIR observations reveal hot and ionised
gas that biconically follow the path of the radio jet
\citep[e.g.,][]{mueller09,ponc08,tomo06,galli04,bock00} and indicate
the existence of a pc-scale warm dust torus \citep[][]{jaffe04}. Early
high-angular resolution radio-continuum observations indicated an
interaction of the radio-jet with the neighbouring insterstellar
medium (ISM) due to the apparently disrupted structure of the
(northern) jet \citep[radio components NE \& C in, e.g.,
][]{gall96,roy98}.  However, in a later publication, \cite{gall04}
argue that the observed disturbed jet-structure (in NE) can as well be
explained by variable outflow speeds due to variable accretion
\citep[e.g.,][]{gall01,siem97} though an earlier collision at position
C was not discarded

The complex dynamics of the molecular gas as traced by the
$^{12}$CO(2--1) line emission within the CND were interpreted as a
consequence of a warped disk \citep{schin00}. However, more recent
observations of the MIR rovibrational H$_2$ emission start to raise
doubts about this interpretation \citep[e.g.,][]{mueller09} and
alternatively suggest that the complex gas kinematics are due to a
funneling of the gas toward the AGN along the jet (inner 60~pc) plus
an expanding ring (on scales of r=100-150~pc), the latter having been
already proposed a few years before by \cite{galli02}.

Additional fascinating characteristics of the CND of NGC~1068, besides
the complex kinematic behaviour of its molecular gas, are its
chemistry and excitation conditions which appear to significantly
differ from starburst/star-forming environments. Traced by the
``abnormal'' line ratios of different molecules and transitions,
mostly by HCN, HCO$^+$ and $^{12}$CO, it has been suggested that the
CND in NGC~1068 harbors a giant X-ray-dominated-region \citep[XDR;
e.g.,][]{kohno08,user04,tac94,ster94}. XDRs are defined in a similar
way to the Photon-Dominated-Regions (PDRs) in starburst galaxies
\citep[such as M82; e.g.,][]{fuen05} but are driven by X-ray rather
than UV-radiation. High HCN-to-CO(J=1--0) ($\geq$1) and
HCN-to-HCO$^+$(J=1--0) ($\geq$1) ratios are found in the CND of
NGC~1068, indicating enhanced HCN abundances there. The X-ray
radiation of the AGN is thereby supposed to be the main driver for the
enhancement of HCN. It can penetrate much deeper into the surrounding
molecular gas than the UV-radiation in PDRs leading to a stimulated
``hyper''-production of HCN.  A multi-transition, multi-molecular line
study of HCN and HCO$^+$ conducted with the IRAM 30m-telescope
\citep[][]{krips08} supports an increased abundance of HCN and/or
increased kinetic temperatures. Both can equally explain the elevated
HCN-to-$^{12}$CO(1--0) line ratios either by the aforementioned
enhancement of the HCN abundance and/or a hypo-excitation of the low-J
$^{12}$CO transitions \citep[see also M51 and NGC~6951 as examples of
increased HCN/$^{12}$CO(J=1--0)
ratios;][]{krip09,krip07,mats07,mats04,kohno96}.

Recent SiO interferometric observations of the CND in NGC~1068 carried
out by \cite{garc10} testify further to the complexity of the gas
chemistry in this galaxy. The bright SiO emission in its CND suggests
an enhanced abundance of this molecule which is interpreted by the
authors as closely related to (high-velocity) shocks. The shocks are
believed to be a consequence of a jet-gas interaction.

\section{Observations}
\label{obs}

A summary of all (sub-)millimeter interferometric observations is
given in Table~\ref{tab2}, where observing parameters as well as
achieved rms noise and angular resolutions are listed.

\subsection{SMA}
For all SMA observations, unless otherwise stated, the phase reference
centre has been set to $\alpha_{\rm J2000}=$02h42m40.70s and
$\delta_{\rm J2000}=$$-$00$^\circ$00$'$$47\farcs9$ which corresponds
to the radio position of the active nucleus
\citep[e.g.][]{gall04,krip06}. The SMA receivers have been tuned to
the respective line using doppler-tracking on the systemic velocity of
NGC~1068 of v$_{\rm LSR}$=1137~\kms. Both the upper sideband (USB) and
lower sideband (LSB) were used for the observations, yielding a
bandwidth of 2~GHz each separated by 10~GHz. A spectral resolution of
0.81~MHz was used for all observations, corresponding to
0.8~km~s$^{-1}$ at 1~mm. The 225~GHz zenith opacity $\tau_{\rm 225}$
was measured regularly throughout all observations at the nearby
Caltech Submillimeter Observatory (CSO). The accuracy of the flux
calibration for all tracks is estimated to be at a conservative level
of $\sim$20\%.

The SMA data have been reduced with SMA specific tasks in the MIR
package \citep[][]{scov93}. Further image analysis has been conducted
with the GILDAS package \citep[][]{guil00}.

\subsubsection{HCN(J=3--2) emission} 
We observed the HCN(J=3--2) line emission in NGC~1068 using the
extended and very extended configurations with up to eight 6~m dishes
in January, October and November 2006.  The 345~GHz receivers were
tuned to the HCN(J=3--2) line (265.886~GHz at rest) in the LSB; the
USB was used for continuum measurements. The weather conditions were
good with opacities of $\tau_{\rm 225}$=0.05-0.1 in the January track
(extended configuration) and $\tau_{\rm 225}$=0.13-0.22 in the
October/November tracks (very extended configuration). We used 3C273,
3C111 and/or 3C454.3 as bandpass and Uranus, Titan, and/or Neptune as
flux calibrators\footnote{We used the line-free USB respectively for
Titan and Neptune to determine the absolute flux level as they are
known to have broad HCN lines which could contaminate a flux
calibration in the LSB.}. We observed two quasars (0235+164, 0238-084,
0339-017, and/or 0423-013) every $\sim$15 minutes to calibrate the
gains (amplitude and phase versus time).  The data from all three
tracks have been combined into one single data file, resulting in an
rms noise of 24~mJy in 17~\kms\ wide velocity channels. The
synthesized beam is determined to be 1$\farcs$0$\times$0$\farcs$8 at a
position angle of PA=30$^\circ$ (natural weighting) and
0$\farcs$53$\times$0$\farcs$46 at PA=30$^\circ$ (robust weighting).

\subsubsection{HCO$^+$(J=3--2) emission}
We carried out observations of the HCO$^+$(J=3--2) emission in
NGC~1068 using seven antennas in extended configuration during
November 2006.  The 345~GHz receivers were tuned to the HCO$^+$(3--2)
line (267.558~GHz at rest) in the LSB; the USB was used for continuum
measurements. The weather conditions were good with opacities of
$\tau_{\rm 225}$=0.06-0.15. Bandpass calibration was performed on
3C273, Titan and Uranus, while absolute fluxes were determined using
Titan. The gains have been calibrated on 0423-013 and 0339-017. For
this data set, we reach an rms noise of 33~mJy in 17~\kms\ wide
velocity channels. The synthesized beam is determined to be
1$\farcs$0$\times$0$\farcs$8 at PA=30$^\circ$ (natural weighting).

\subsubsection{$^{12}$CO(J=3--2) and HCO$^+$(J=4--3) emission} 
The $^{12}$CO(J=3--2) emission of NGC~1068 was observed in extended
configuration using all eight antennas during September
2007. HCO$^+$(J=4--3) was additionally observed in a separate track in
August 2007. The 345~GHz receivers were tuned to the $^{12}$CO(J=3--2)
line (345.796~GHz at rest) in the LSB such that the HCO$^+$(J=4--3)
(356.734~GHz at rest) still falls within the USB. The opacities ranged
between $\tau_{\rm 225}$=0.06-0.13. Bandpass calibration was performed
on 3C454.3 and Uranus. Uranus was also used for flux
calibration. Gains were determined using 0238+166 and 0423-013. An rms
noise of 80~mJy is reached in 7~\kms\ wide velocity channels. The
synthesized beam is determined to be 1$\farcs$0$\times$0$\farcs$8 at
PA=30$^\circ$ for natural weighting when also using a uv-taper to
better match the angular resolutions of the other SMA observations.
The original (un-tapered) angular resolution amounts to 0.6$\times$0.5
at PA=30$^\circ$.

\subsubsection{$^{13}$CO(J=2--1) and C$^{18}$O(J=2--1) emission}
The $^{13}$CO(J=2--1) emission in NGC~1068 has been observed in
extended configuration using all eight antennas during January and
February 2008. The 230~GHz receivers have been tuned to the
$^{13}$CO(J=2--1) line (220.399~GHz at rest) in the USB such that the
C$^{18}$O(J=2--1) (219.560~GHz at rest), the HC$_3$N(J=23-22)
(209.230~GHz at rest) and H32$\alpha$ (210.502~GHz at rest) still fall
within the LSB. However, only the $^{13}$CO(J=2--1) and
C$^{18}$O(J=2--1) line emission was detected.  The opacities ranged
between $\tau_{\rm 225}$=0.1-0.2. Bandpass calibration was performed
on 0423-013, 3C111 and Titan, while gains were determined using
0339-017 and 0423-013. Titan was further used as a flux calibrator. An
rms noise of 12~mJy was reached in 17~\kms\ wide velocity
channels. The synthesized beam is determined to be
1$\farcs$0$\times$0$\farcs$9 at PA=30$^\circ$ for natural weighting
when also using a uv-taper to better match the angular resolutions of
the other SMA observations.

\subsubsection{$^{13}$CO(J=3--2) emission}
The $^{13}$CO(J=3--2) emission in NGC~1068 was observed in compact
configuration using seven antennas in October 2005. These observations
were part of the observing campain presented by \cite{hump05}, which
aimed to detect extragalactic H$_2$O maser emission at
(sub-)millimeter wavelengths. The 345~GHz receivers have been tuned to
the H$_2$O(10(2,9)-9(3,6)) maser line (321.226~GHz at rest) in the LSB
such that the $^{13}$CO(J=3--2) line (356.734~GHz at rest) was still
located within the USB. The 225~GHz zenith opacity has remained stable
around 0.05-0.06. Bandpass calibration has been performed on 3C454.3,
3C111 and Uranus. Uranus has been also used as a flux calibrator. The
gains were determined using 0234+285 and verified against 0215+015 and
0420-014. An rms noise of 61~mJy is reached in 7~\kms\ wide velocity
channels. The synthesized beam is determined to be
2$\farcs$4$\times$2$\farcs$1 at PA=28$^\circ$ (natural weighting).

\subsection{IRAM PdBI}

\subsubsection{$^{12}$CO(2--1) and $^{13}$CO(J=1--0) emission}
Observations of the $^{12}$CO(2--1) emission in NGC~1068 were carried
out with the IRAM PdBI in February 2003 using all six antennas in A
configuration. Simultaneously, we observed the $^{13}$CO(J=1--0) using
the 3~mm PdBI receivers.  The bandpass was calibrated on NRAO150 and
0420-014 while phase and amplitude calibration were performed on
0235+164 and 0238-084. A total bandwidth of 580~MHz with a spectral
resolution of 1.25~MHz was used. We reach an RMS of $\sim$7~mJy in
7~\kms\ wide channels (natural weighting) at 1~mm and of $\sim$1.9~mJy
in 14~\kms\ wide channels (natural weighting) at 3~mm. Applying
natural weighting in the mapping process, beam sizes are derived to be
1$\farcs$0$\times$0$\farcs$6 at PA=36$^\circ$ at 1~mm and
2$\farcs$5$\times$1$\farcs$7 at PA=28$^\circ$ at 3~mm. However, to
better match the SMA observations, we mapped the $^{12}$CO(2--1) data
with a uv-taper giving an effective angular resolution of
1.0$''\times$0.8$''$ at PA=30$^\circ$. As the uv-coverages between the
SMA and PdBI data (for the high angular resolution data) are very
similar, the usage of a simple uv-taper already provides the necessary
accuracy to match the restoring beam of the PdBI observations with
that of the SMA.

\section{Results}
\label{res}

\subsection{Continuum emission: from 850~$\mu$m to 1.4~mm}
The sub(mm) continuum emission at the wavelengths presented in this
paper was derived from the line-free channels of the respective line
observations (Table~\ref{tab2}), averaging emission from USB and LSB
where possible (for the SMA data; the PdBI data were obtained from
single-sideband observations).  The continuum emission has further
been merged between data sets with very similar observed frequencies
(i.e., within $\sim$10~GHz). Before averaging data from different
sidebands and/or observations, it has been carefully verified that the
absolute flux, position and structure of the emission in the
individual data sets are consistent with each other within the
calibrational uncertainties of 20\% (flux) and ~0.1$''$ (position) in
order to reduce systematic errors.

The continuum emission is clearly detected at all wavelengths at a
$\geq$5$\sigma$ level.  To obtain accurate fluxes, positions and
sizes, elliptical Gaussians were fitted to the data in the uv-plane,
except for the uniformly weighted map of the 1.0~mm continuum emission
for which a circular Gaussian was fitted given its apparently
unresolved nature. The results of these fits are listed in
Table~\ref{tab3}.

\begin{figure*}[!]
\centering
\rotatebox{-90}{\resizebox{!}{\hsize}{\includegraphics{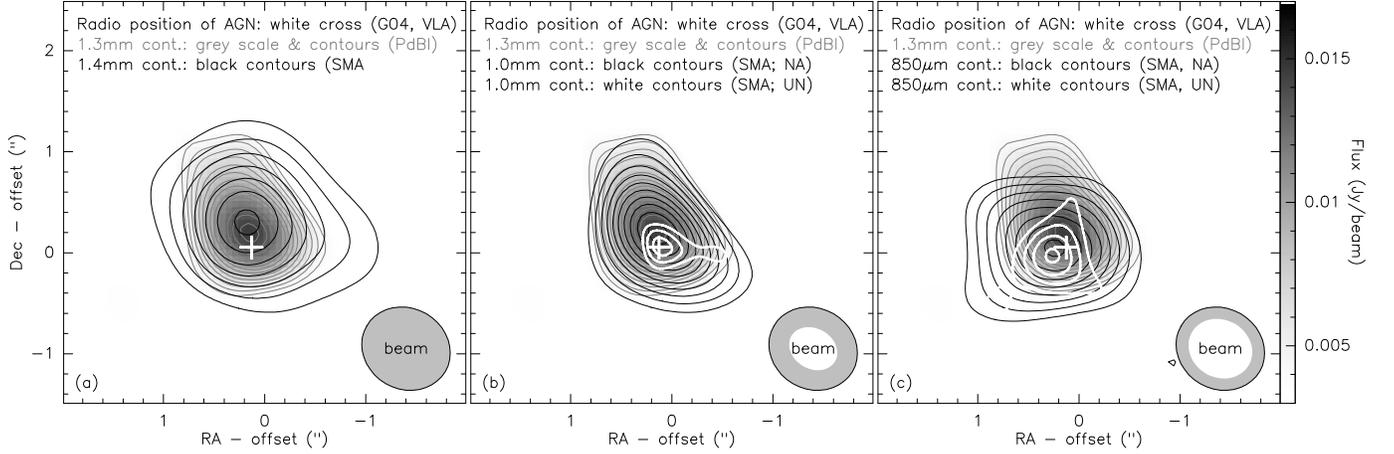}}}
\caption{Continuum emission of NGC~1068 at $\lambda$=1.4~mm (Fig.~1a;
  black contours), 1.3~mm (Fig.~1a-c; grey scale and grey contours),
  1.0~mm (Fig.~1b; black contours) and 850~$\mu$m (Fig.~1c; black
  contours), observed with the SMA and the IRAM PdBI. The white cross
  denotes the position of the AGN measured by \citeauthor{gall04}
  \citeyearpar[G04]{gall04}. The contours of the 1.3~mm continuum
  emission (PdBI) start at 5$\sigma$=4~mJy in steps of 1$\sigma$. {\it
  a)} The contours of the 1.4~mm continuum emission (SMA) start at
  5$\sigma$=4~mJy in steps of 1$\sigma$. {\it b)} The contours of the
  1.0~mm continuum emission (SMA, NA) start at 3$\sigma$=1.6~mJy in
  steps of 1$\sigma$, while the contours of the uniformaly mapped
  1.0~mm continuum emission run from 3$\sigma$=2.3~mJy in steps of
  1$\sigma$. {\it c)} The contours of the 850~$\mu$m continuum
  emission (SMA, NA) start at 3$\sigma$=2.4~mJy in steps of 1$\sigma$,
  while the contours of the uniformaly mapped 850~$\mu$m continuum
  emission run from 3$\sigma$=2.6~mJy in steps of 1$\sigma$. }
\label{fig1.0}
\end{figure*}

The continuum emission at 1.0~mm (NA), 1.3~mm and 1.4~mm appear to be
consistent with each other in terms of their flux, position and
structure (see Table~\ref{tab3} and Fig.~\ref{fig1.0}; see also
\citeauthor{krip06} \citeyear{krip06}).  All show peak fluxes of
around 15-19~mJy/beam and spatially integrated flux densities of
22-28~mJy, indicating extended emission.  Their positions, although
self-consistent, are slightly to the North ($\sim$0.2$''$) of the
radio position of the AGN \citep[component S1 from][marked with a
white cross in Fig.~\ref{fig1.0}]{gall04} and that of the uniformly
weighted 1.0~mm continuum emission (Fig.~\ref{fig1.0}b; white
contours).  The shift between the mm and cm data is larger than the
positional uncertainty of 0.1$''$ and thus assumed to be real.
NGC~1068 is known to have a pronounced radio (and mm-) jet in a
North-East-to-South-West direction, of which the North-Eastern part
exhibits the stronger emission \citep[e.g.,][]{gall04,krip06}.
Despite the steep spectral index of the synchrotron emission of the
jet, the extended (i.e., $>$1$''$) emission from both the jet and
counter-jet are still visible at 3~mm \citep[e.g.,][]{krip06,schin00}
but are significantly fainter or undetected at shorter wavelengths
(i.e., $\lesssim$1~mm).

The continuum emission at 1.0~mm (NA), 1.3~mm and 1.4~mm is a blend of
emission associated with the (North-East) radio jet and the AGN itself
\citep[S1 in][]{gall04} due to the 'lower' angular resolution of
$\sim$1$''$. Due to the higher angular resolution, the 1.0~mm
continuum emission of the jet in the uniformly weighted map
(Fig.~\ref{fig1.0}b) is almost entirely resolved out leaving behind
only the more compact emission of the AGN. Thus, the centroid of the
emission at 1.0~mm (NA), 1.3~mm and 1.4~mm will naturally be shifted
towards the North, while the 1.0~mm (UN) continuum emission should
reveal the actual position of the AGN (or at least the base of the
jet).

Going to even shorter wavelengths of 850~$\mu$m, it appears that not
only the continuum flux increases again, but also its position seems
to be now consistent with the AGN, independent of the weighting (i.e.,
synthesized beam) used for mapping/cleaning and unlike the 1.0~mm
(NA), 1.3mm, and 1.4~mm continuum emission. The latter may indicate
that the emission from the radio jet is negligeable at 850~$\mu$m (see
also Fig.~3 in \citeauthor{krip06} \citeyear{krip06}) and the AGN
(i.e., the S1 component) dominates. The increased flux at 850~$\mu$m,
which appears to be larger by almost a factor of 2 compared to the
1.0~mm-1.4~mm emission, shows that thermal dust emission already plays
a significant role at 850~$\mu$m (see Section~\ref{sed}). Also, the
size and shape of the continuum emission appear to have changed
compared to that at longer wavelengths. The PA of the 850$\mu$m
emission of $\sim$90$^\circ$ is significantly different from that
($\sim$30$^\circ$) of the 1.0~mm-1.4~mm emission. Moreover, the
850~$\mu$m emission appears to be extended (Fig.~\ref{fig1.0}c), in
contrast to the 1.0~mm-1.4~mm emission, which seems to be extended
only in the jet component but not in the 'left-over' AGN component in
the uniformly weighted 1.0~mm map (Fig.~\ref{fig1.0}b). The uniformly
weighted 850~$\mu$m continuum emission appears to be also resolved
(white contours in Fig.~\ref{fig1.0}c, compare also peak flux with
total flux density in Table~\ref{tab3}).

\subsection{Line emission}

\subsubsection{General Characteristics and Distribution of the Molecular Gas}
\begin{figure*}[!]
\centering
\rotatebox{0}{\resizebox{!}{\hsize}{\includegraphics{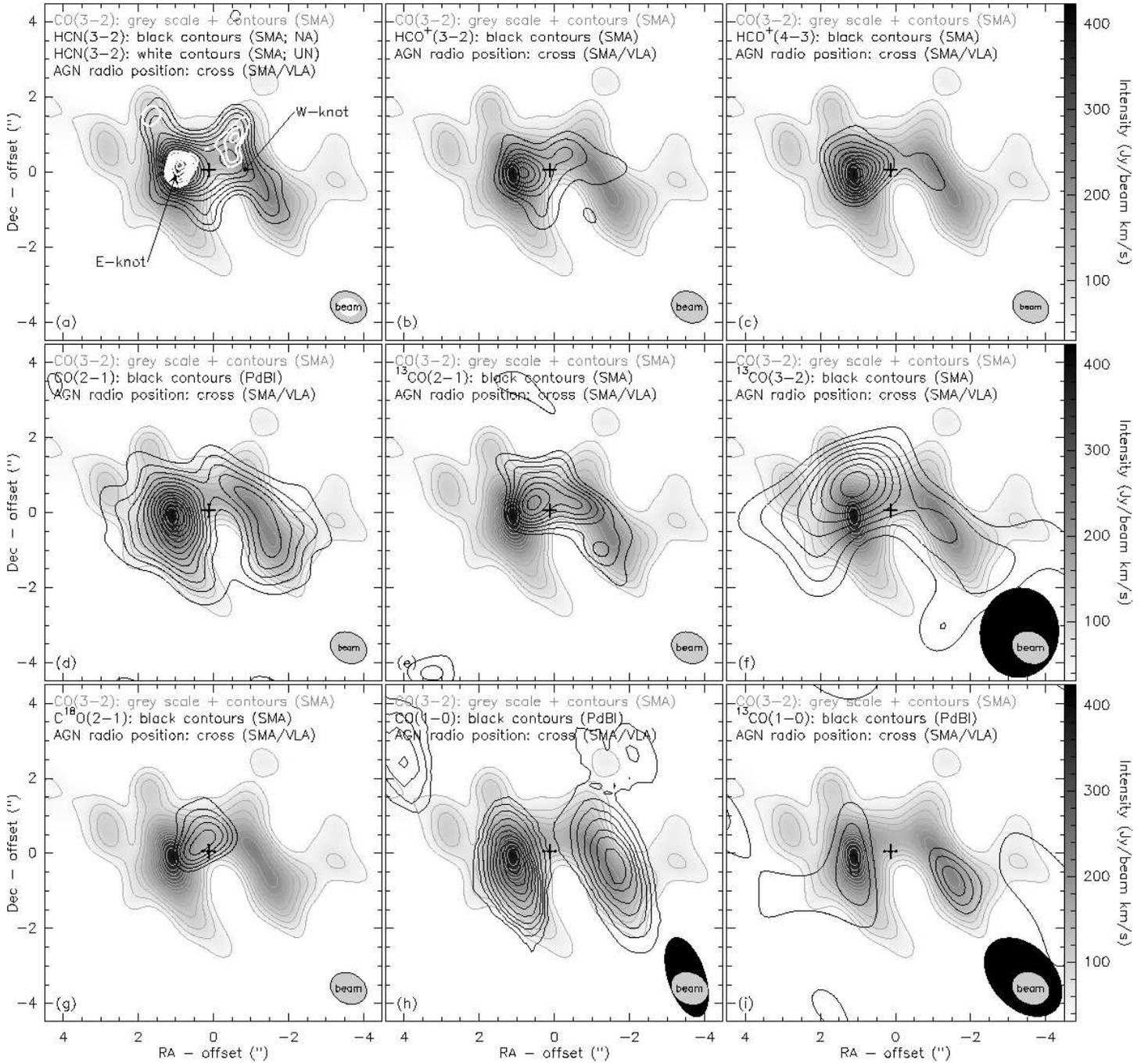}}}
\caption{Velocity integrated line emission of $^{12}$CO(J=3--2) ({\it
    a-g}), HCN(J=3--2) ({\it a}), HCO$^+$(J=3--2) ({\it b}),
    HCO$^+$(J=4--3) ({\it c}), $^{12}$CO(J=2--1) ({\it d}),
    $^{13}$CO(J=2--1) ({\it e}), $^{13}$CO(J=3--2) ({\it f}),
    C$^{18}$O(J=2--1) ({\it g}), $^{12}$CO(J=1--0) ({\it h}) and
    $^{13}$CO(J=1--0) ({\it i}) in NGC~1068, observed with the SMA and
    the IRAM PdBI. Contour levles are: $^{12}$CO(J=3--2) -- from
    10$\sigma$ by 6$\sigma$ with 1$\sigma$=4.8~Jy~km~s$^{-1}$;
    HCN(J=3--2) -- from 3$\sigma$ by 1$\sigma$ with
    1$\sigma$=2.6~Jy~km~s$^{-1}$; HCO$^+$(J=3--2) -- from 2$\sigma$ by
    1$\sigma$ with 1$\sigma$=3.4~Jy~km~s$^{-1}$; HCO$^+$(J=4--3) --
    from 3$\sigma$ by 1$\sigma$ with 1$\sigma$=2.4~Jy~km~s$^{-1}$;
    $^{12}$CO(J=2--1) -- from 5$\sigma$ by 5$\sigma$ with
    1$\sigma$=1.2~Jy~km~s$^{-1}$; $^{13}$CO(J=2--1) -- from 3$\sigma$
    by 1$\sigma$ with 1$\sigma$=0.9~Jy~km~s$^{-1}$; $^{13}$CO(J=3--2)
    -- from 5$\sigma$ by 1$\sigma$ with 1$\sigma$=3.5~Jy~km~s$^{-1}$;
    C$^{18}$O(J=2--1) -- from 3$\sigma$ by 1$\sigma$ with
    1$\sigma$=0.9~Jy~km~s$^{-1}$; $^{12}$CO(J=1--0) -- from 5$\sigma$
    by 3$\sigma$ with 1$\sigma$=0.4~Jy~km~s$^{-1}$; $^{13}$CO(J=1--0)
    -- from 1$\sigma$ by 1$\sigma$ with 1$\sigma$=0.1~Jy~km~s$^{-1}$.}
\label{fig2.0}
\end{figure*}

The continuum emission has been subtracted from all line data in the
uv plane to avoid any contamination of the line by continuum emission
even if in some cases the continuum emission does not exceed the noise
level in the individual channel maps (see Table~\ref{tab2} \&
\ref{tab3}). In order to reduce systematic effects due to spatial
filtering, we used a slight uv-taper (giving some more weight to the
shorter baselines) to map and clean all line emission data with the
same synthesized beam, except for the high angular resolution
($\sim$0.4-0.5$''$) of the uniformly weighted HCN(J=3--2) map (shown
additionally in Fig.~\ref{fig2.0}a using white contours) and the low
angular resolution ($\sim$2$''$) of the $^{12}$CO(J=1--0),
$^{13}$CO(J=1--0) and $^{13}$CO(J=3--2) maps.

Fig.~\ref{fig2.0} shows the velocity integrated intensity maps of the
molecular line emission from HCN(J=3--2), HCO$^+$(J=3--2),
HCO$^+$(J=4--3), $^{12}$CO(J=2--1), $^{13}$CO(J=2--1),
$^{13}$CO(J=3--2), C$^{18}$O(J=2--1), $^{12}$CO(J=1--0) and
$^{13}$CO(J=1--0); $^{12}$CO(J=3--2) is plotted in grey scale in all
images to facilitate a comparison. All these molecules have been
clearly detected above the 5$\sigma$ level (except
$^{13}$CO(J=1--0)). The emission in all lines reveals a pronounced
peak on the stronger eastern knot and in the stronger lines also a
weaker peak on the western knot, both of which are already known from
previous $^{12}$CO observations \citep[e.g.,][]{schin00}.  An
elliptical Gaussian has been fitted to the uv-data for all lines in
order to obtain the position, peak- and spatially integrated flux of
the emission in the two knots. The results of the fits are given in
Table~\ref{tab4}. The position of the emission in the eastern and
western knot is very similar in all observed lines, excluding the
$^{13}$CO(J=2--1) and C$^{18}$O(J=2--1) emission which seem to peak
closer to the AGN.

\begin{figure*}[!]
\centering
\rotatebox{0}{\resizebox{\hsize}{!}{\includegraphics{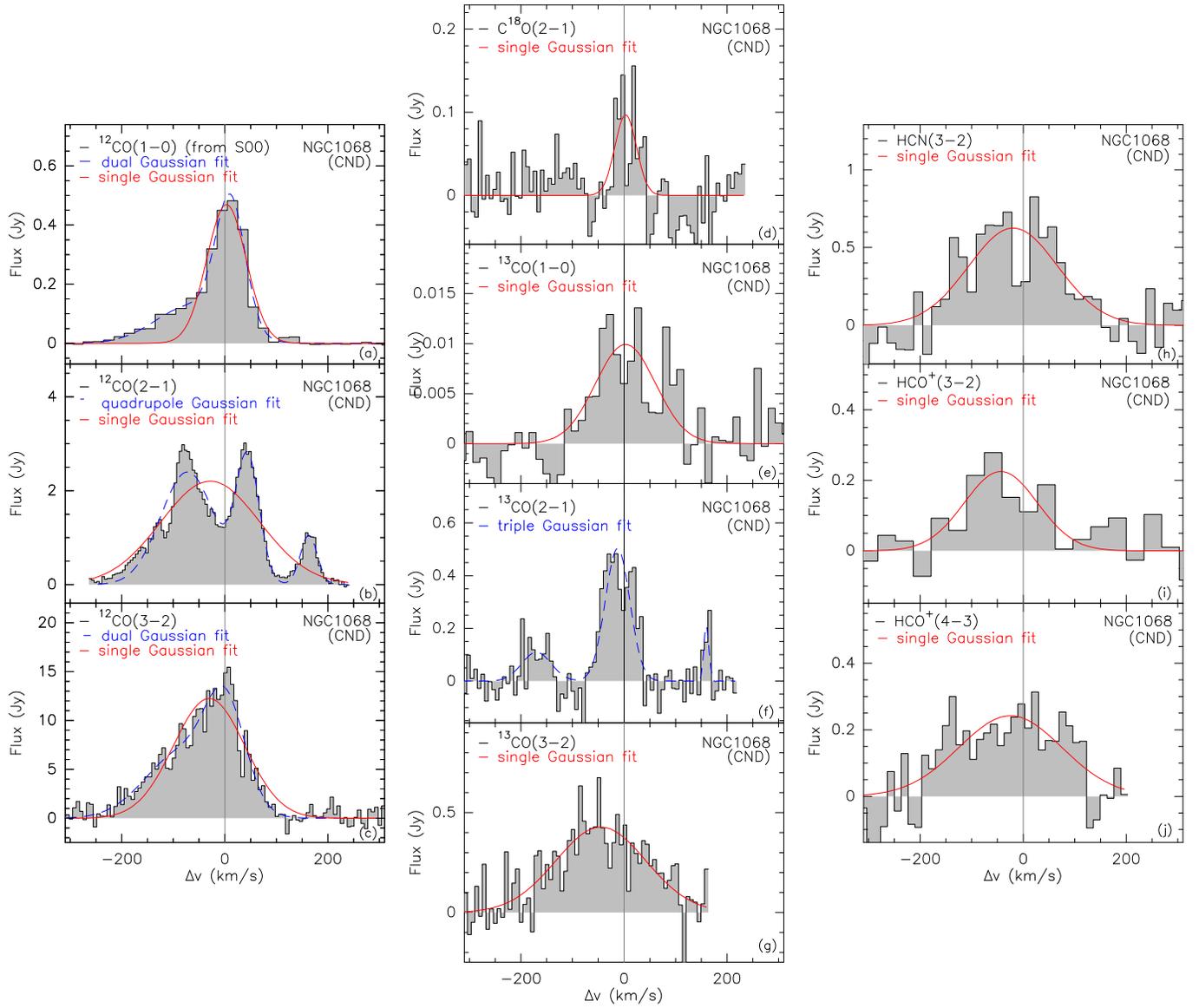}}}
\caption{Spatially integrated spectrum of different molecular lines in
  NGC~1068. The single-line Gaussian fit is indicated with a dotted red
  line (parameters are listed in Table~\ref{tab5}) while the multiple
  Gaussian fit is plotted with a dashed blue line.}
\label{fig3.0}
\end{figure*}

The spectrum of the spatially integrated emission (over the central
$\sim$4$''$) of each line is plotted in Fig.~\ref{fig3.0}. We also
show the $^{12}$CO(J=1--0) line emission taken from \cite{schin00} for
consistency. While the velocity integrated line emission seems to be
very similar in its shape and position for most lines, the line
profiles vary significantly from each other. While for the dense gas
tracers (HCN, HCO$^+$) a single Gaussian fit is sufficient to
reproduce the line profiles, the CO lines need a dual, triple or
quadrupole Gaussian fit. However, to simplify a comparison, the
results given in Table~\ref{tab5} represent a single Gaussian fit to
all lines. The line centers are roughly consistent with each other,
differing by less than 20~kms~s$^{-1}$. Excluding the
$^{13}$CO(J=2--1) and C$^{18}$O(J=2--1) emission, the line widths also
agree with each other within the uncertainties. Except for the
$^{12}$CO(J=2--1) and $^{13}$CO(J=2--1) line emission for which
roughly half of the emission seems to be resolved out, the
interferometric observations have captured most of the emission
measured with single dish observations (Table~\ref{tab5}). Please note
that for the $^{12}$CO(J=1--0) \citep{schin00} and $^{13}$CO(J=1--0)
emission, the single dish fluxes are much higher than the
interferometric ones because they contain significant emission from
the star-forming ring/spiral-arms and the bar.

\subsubsection{Dynamical Characteristics of the Molecular Gas}
\begin{figure}[!]
\centering
\rotatebox{0}{\resizebox{\hsize}{!}{\includegraphics{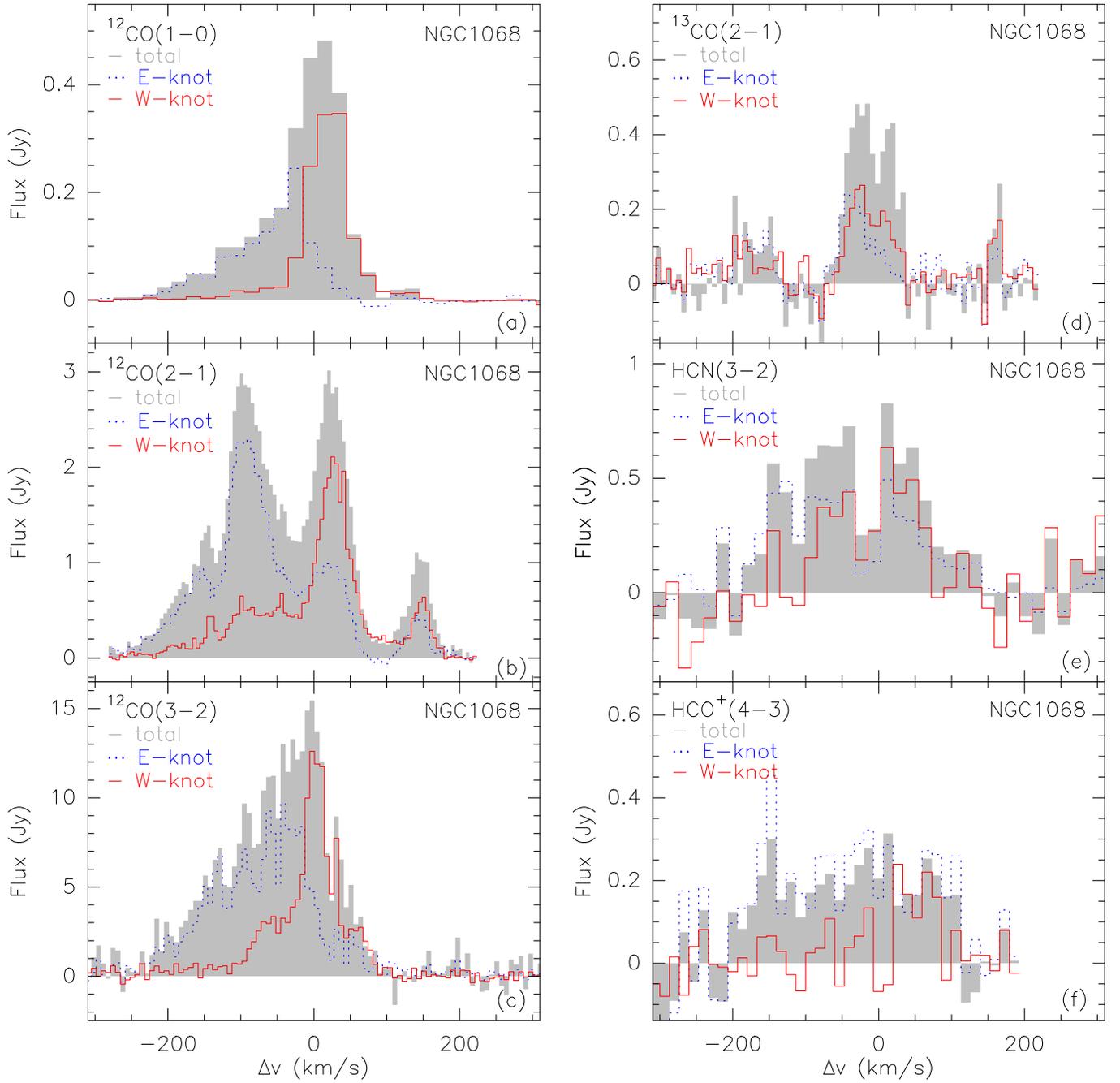}}}
\caption{Spatially integrated spectrum of the $^{12}$CO(J=1--0),
  \citep[taken from ][]{schin00}, $^{12}$CO(J=2--1) and
  $^{12}$CO(J=3--2) ({\it left column}) and HCN(J=3--2) and
  HCO$^+$(J=4--3) emission over the E-knot ({\it dotted blue}) and W-knot
  ({\it solid red}) component of NGC~1068. }
\label{fig4.0}
\end{figure}

\begin{figure*}[!t]
\centering
\rotatebox{0}{\resizebox{\hsize}{!}{\includegraphics{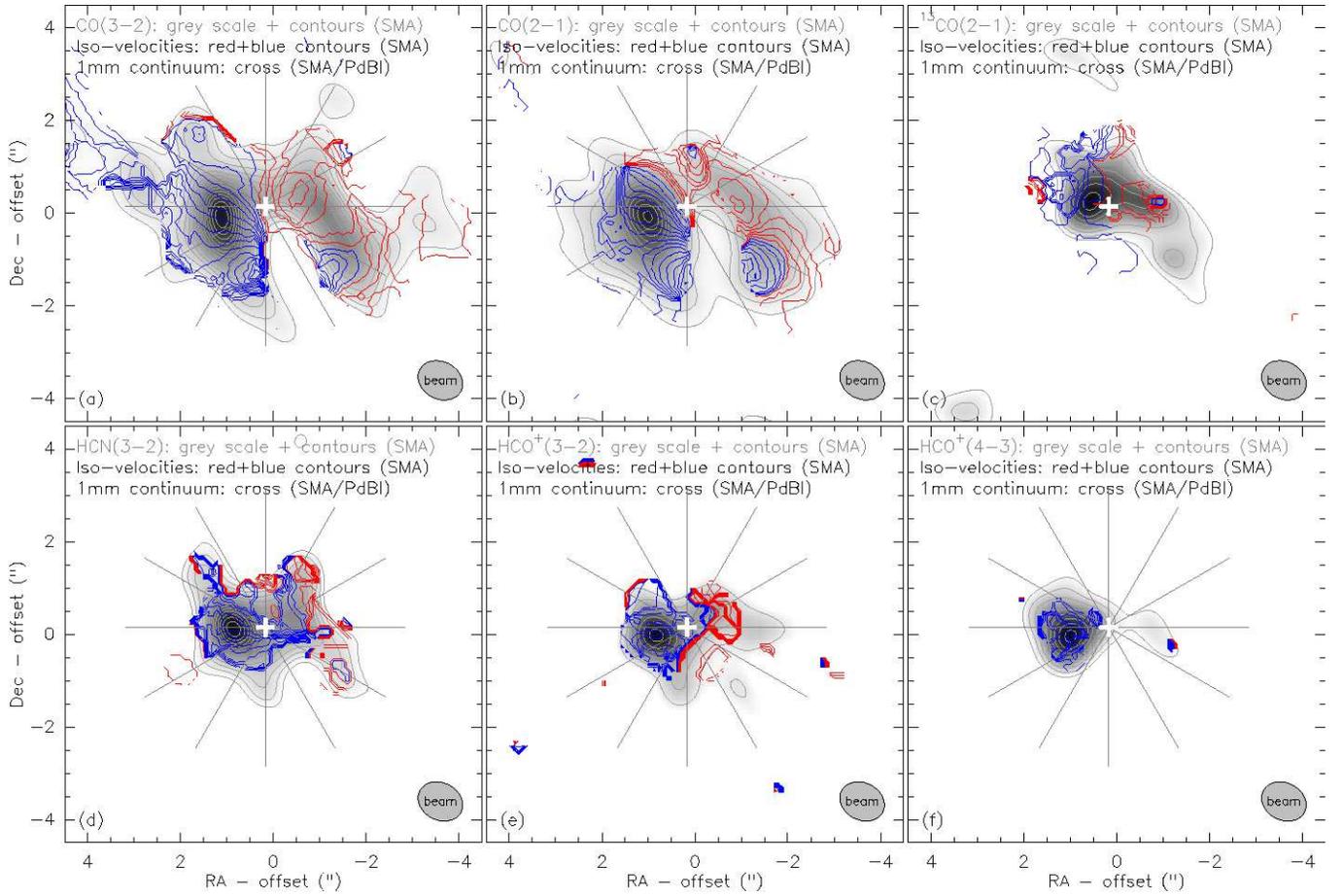}}}
\caption{Iso-velocity maps of the different molecular lines observed
  in NGC~1068. The grey scale correspond to the velocity integrated
  emission of each line with the same contours as used in
  Fig.~\ref{fig2.0}. The velocity contours are in steps of
  10~km~s$^{-1}$ around the systemic velocity of NGC~1068. The grey
  lines indicate the cuts along which the position-velocity diagrams
  (see Fig.~\ref{fig10.0}) were taken for the respective molecules
  ($^{12}$CO(J=2--1),$^{12}$CO(J=2--1) \& HCN(J=3--2)).}
\label{fig5.0}
\end{figure*}

\begin{figure*}[!t]
\centering
\rotatebox{0}{\resizebox{\hsize}{!}{\includegraphics{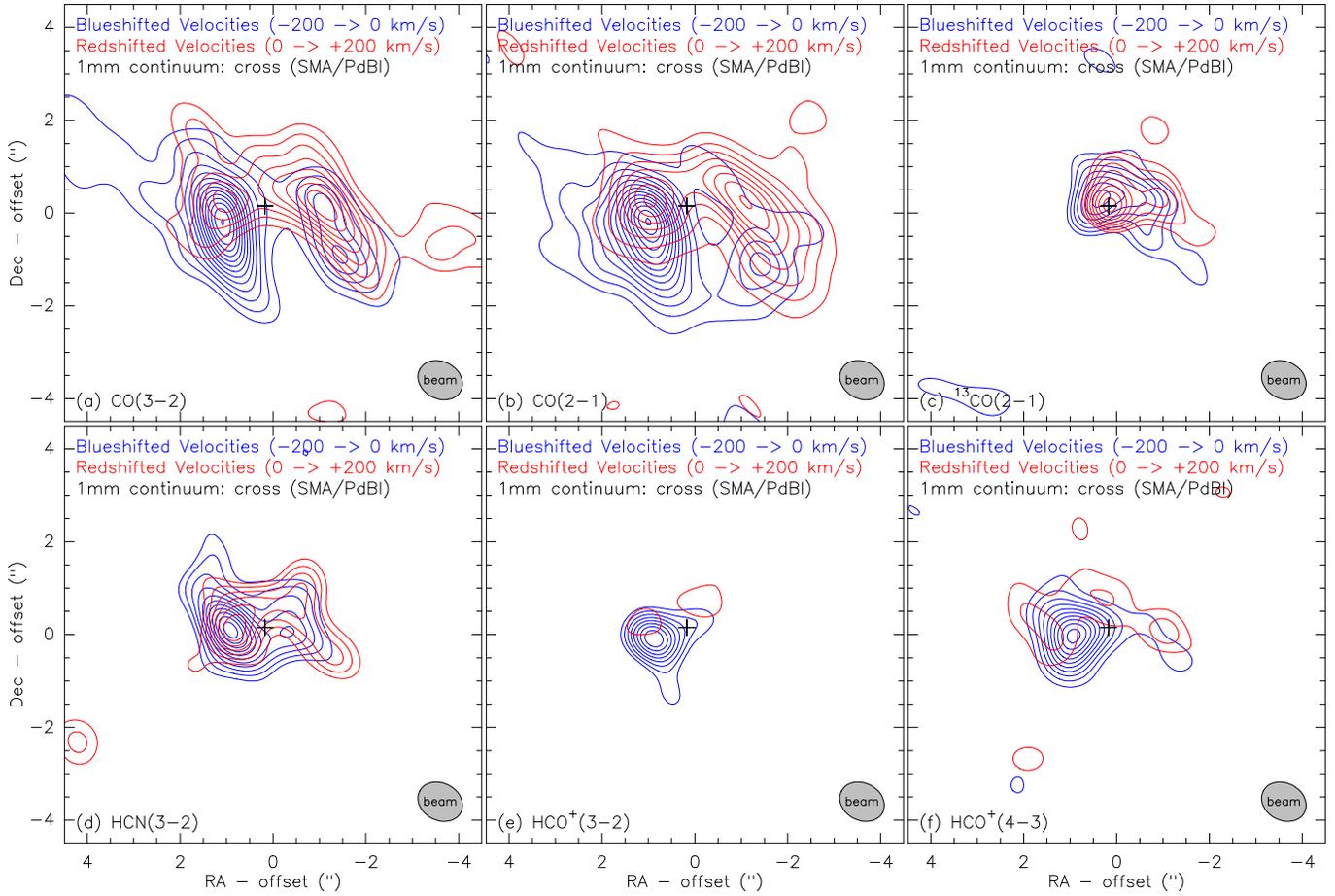}}}
\caption{Red- and blueshifted emission of NGC1068. Contours are in
steps of a) from 7$\sigma$ by 5$\sigma$ with
1$\sigma$=2.0~Jy~km~s$^{-1}$; b) from 5$\sigma$ by 5$\sigma$ with
1$\sigma$=0.9~Jy~km~s$^{-1}$; c) from 4$\sigma$ by 1$\sigma$ with
1$\sigma$=0.5~Jy~km~s$^{-1}$; d) from 3$\sigma$ by 1$\sigma$ with
1$\sigma$=1.9~Jy~km~s$^{-1}$; e) from 3$\sigma$ by 1$\sigma$ with
1$\sigma$=2.3~Jy~km~s$^{-1}$; f) from 2$\sigma$ by 1$\sigma$ with
1$\sigma$=1.4~Jy~km~s$^{-1}$.}
\label{fig6.0}
\end{figure*}

\begin{figure*}[!t]
\centering
\rotatebox{-90}{\resizebox{!}{\hsize}{\includegraphics{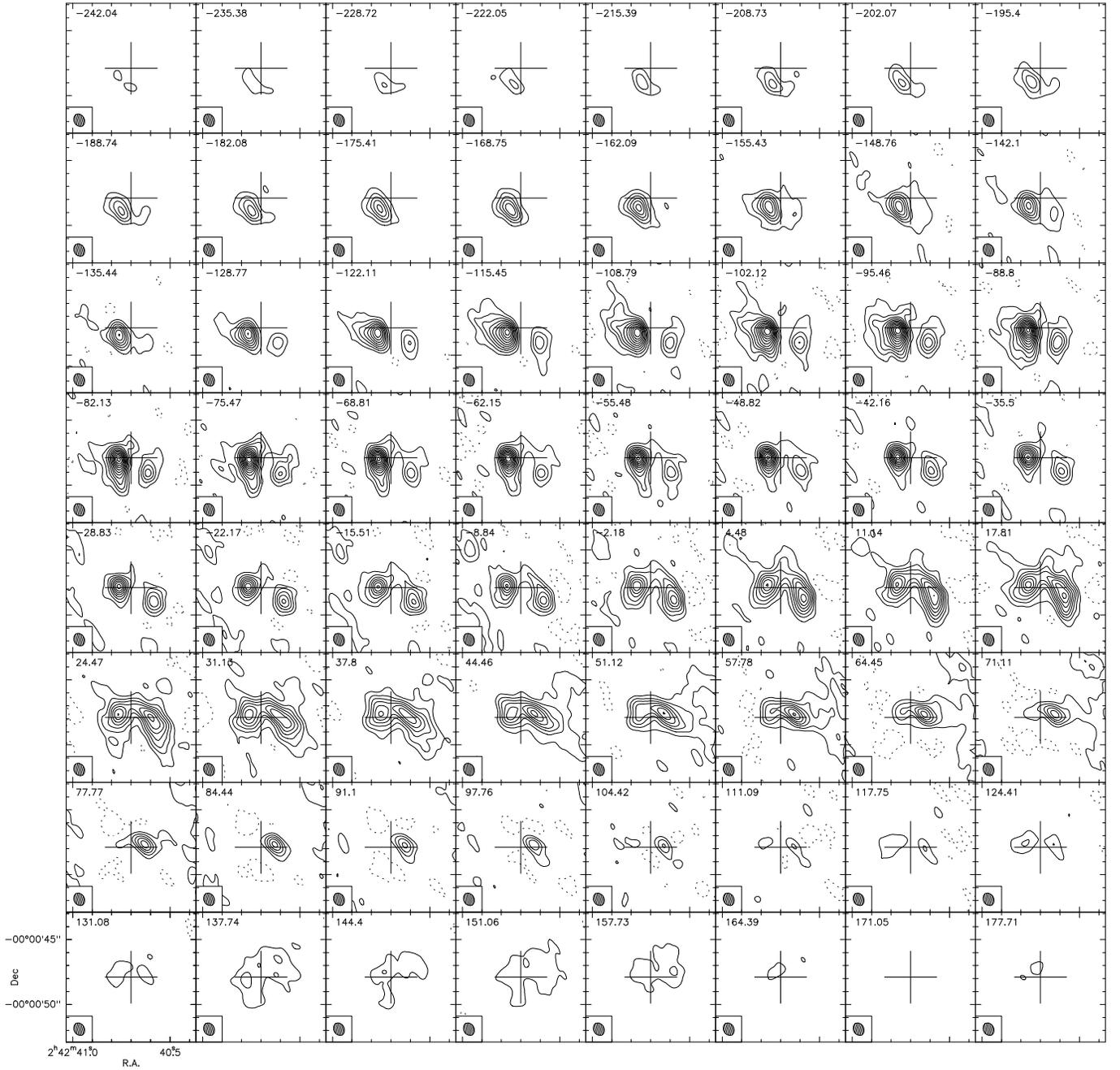}}}
\caption{Channel maps of the $^{12}$CO(J=2--1) emission in
  NGC~1068. Please note that the $^{12}$CO(J=2--1) data were resampled
  to match the spectral resolution of the $^{12}$CO(J=3--2) data and
  facilitate a comparison, especially with respect to the line-ratio
  channel map in Fig.~\ref{fig12.0}. Contour spacing is in steps of
  5$\sigma$=37.4~mJy~beam$^{-1}$. We use a spectral resolution of
  $\sim$7~km/s. The zero channel corresponds to v$_{\rm
  LSR}$=1137~\kms\ of NGC~1068.}
\label{fig7.0}
\end{figure*}

\begin{figure*}[!t]
\centering
\rotatebox{-90}{\resizebox{!}{\hsize}{\includegraphics{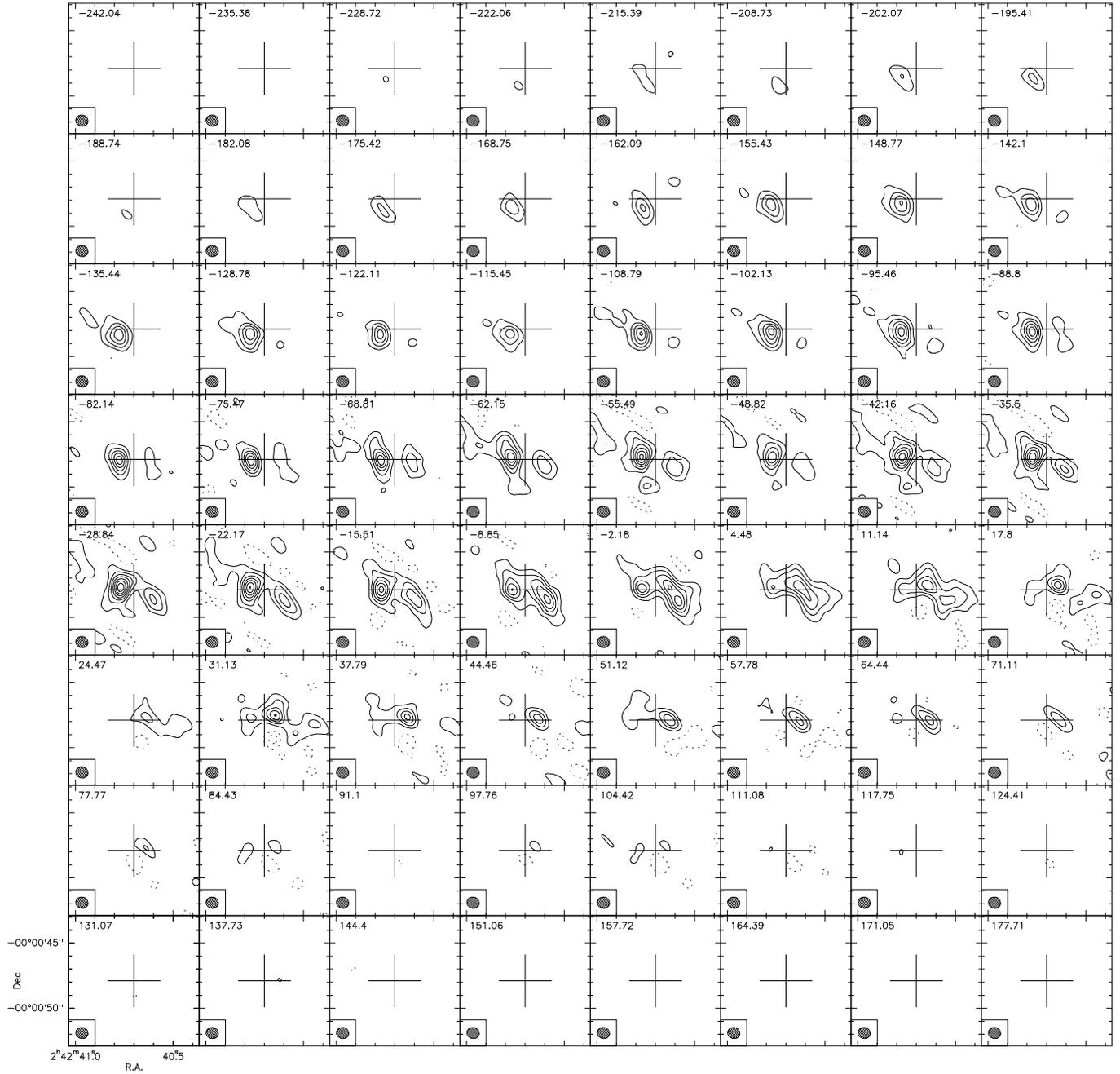}}}
\caption{Channel maps of the $^{12}$CO(J=3--2) emission in
  NGC~1068. We use a spectral resolution of $\sim$7~km~s$^{-1}$ and
  the original spatial resolution from the observations. Contour
  spacing is in steps of 5$\sigma$=400~mJy~beam$^{-1}$. The zero
  channel corresponds to v$_{\rm LSR}$=1137~\kms\ of NGC~1068.}
\label{fig8.0}
\end{figure*}

\begin{figure*}[!t]
\centering
\rotatebox{-90}{\resizebox{!}{\hsize}{\includegraphics{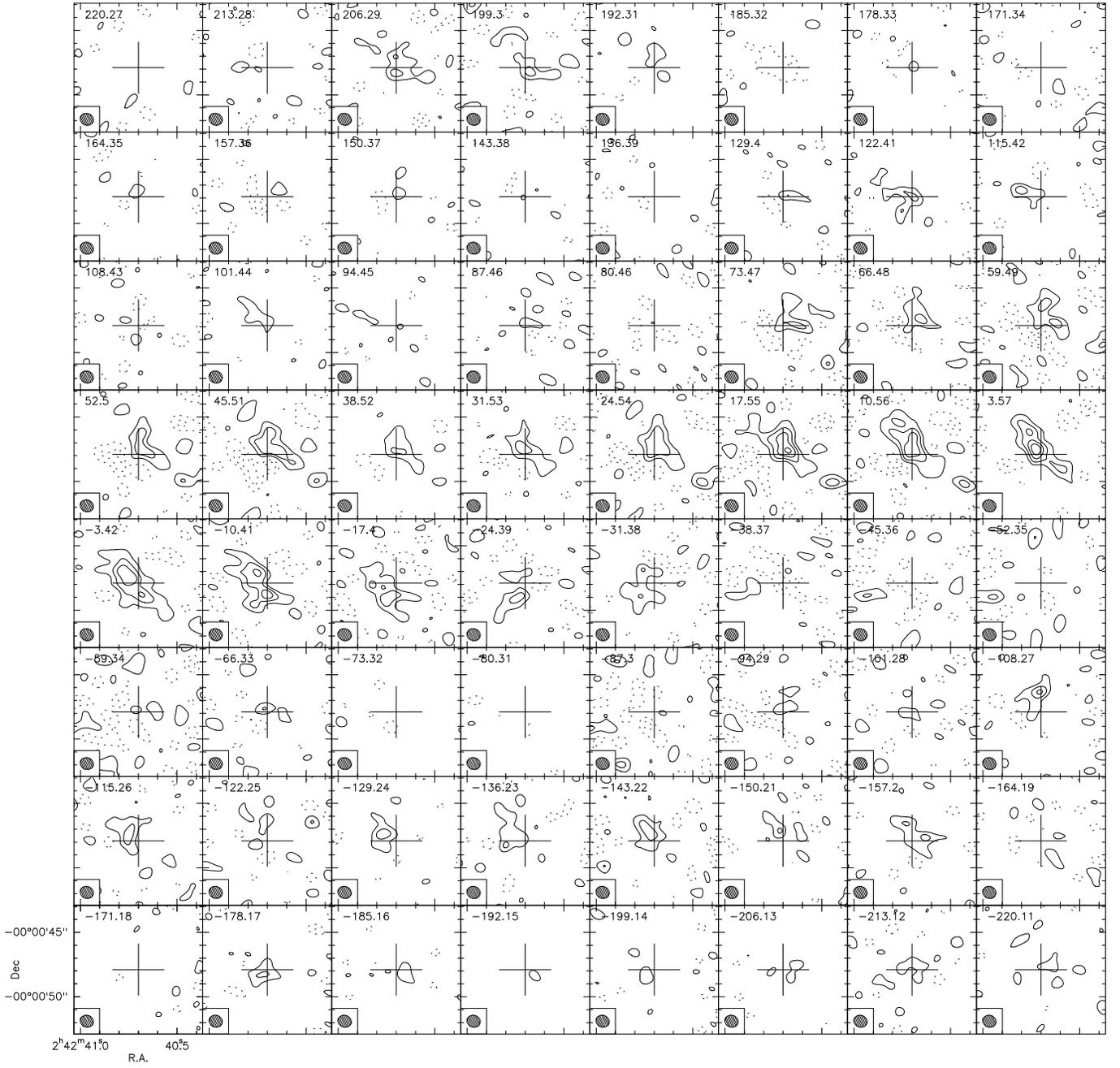}}}
\caption{Channel maps of the $^{13}$CO(J=2--1) emission in
  NGC~1068. We use spectral resolution of $\sim$7~km~s$^{-1}$. Contour
  spacing is in steps of 2$\sigma$=11~mJy~beam$^{-1}$. The zero
  channel corresponds to v$_{\rm LSR}$=1137~\kms\ of NGC~1068. }
\label{fig9.0}
\end{figure*}

\begin{figure*}[!t]
\centering
\rotatebox{0}{\resizebox{\hsize}{!}{\includegraphics{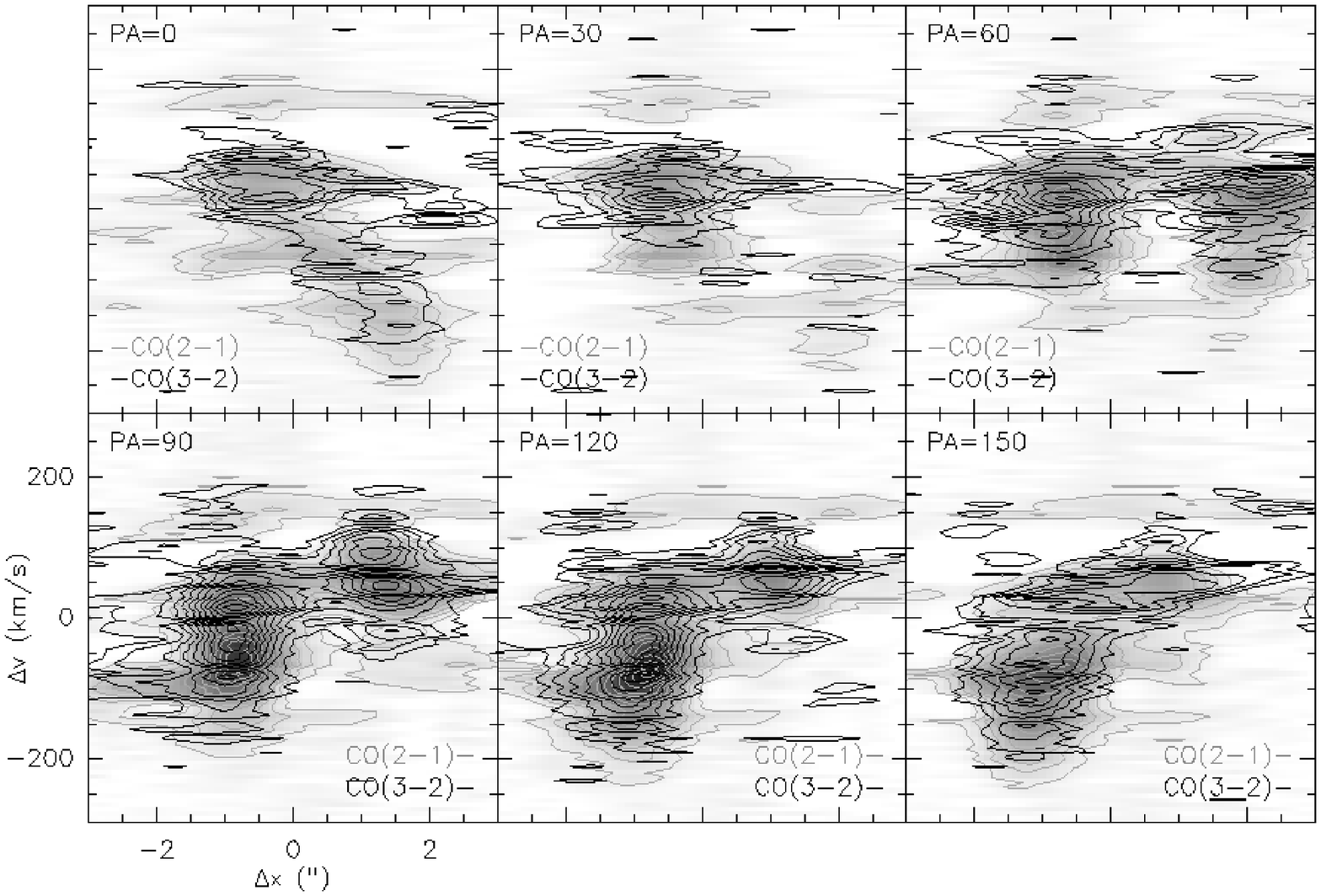}}}
\rotatebox{0}{\resizebox{\hsize}{!}{\includegraphics{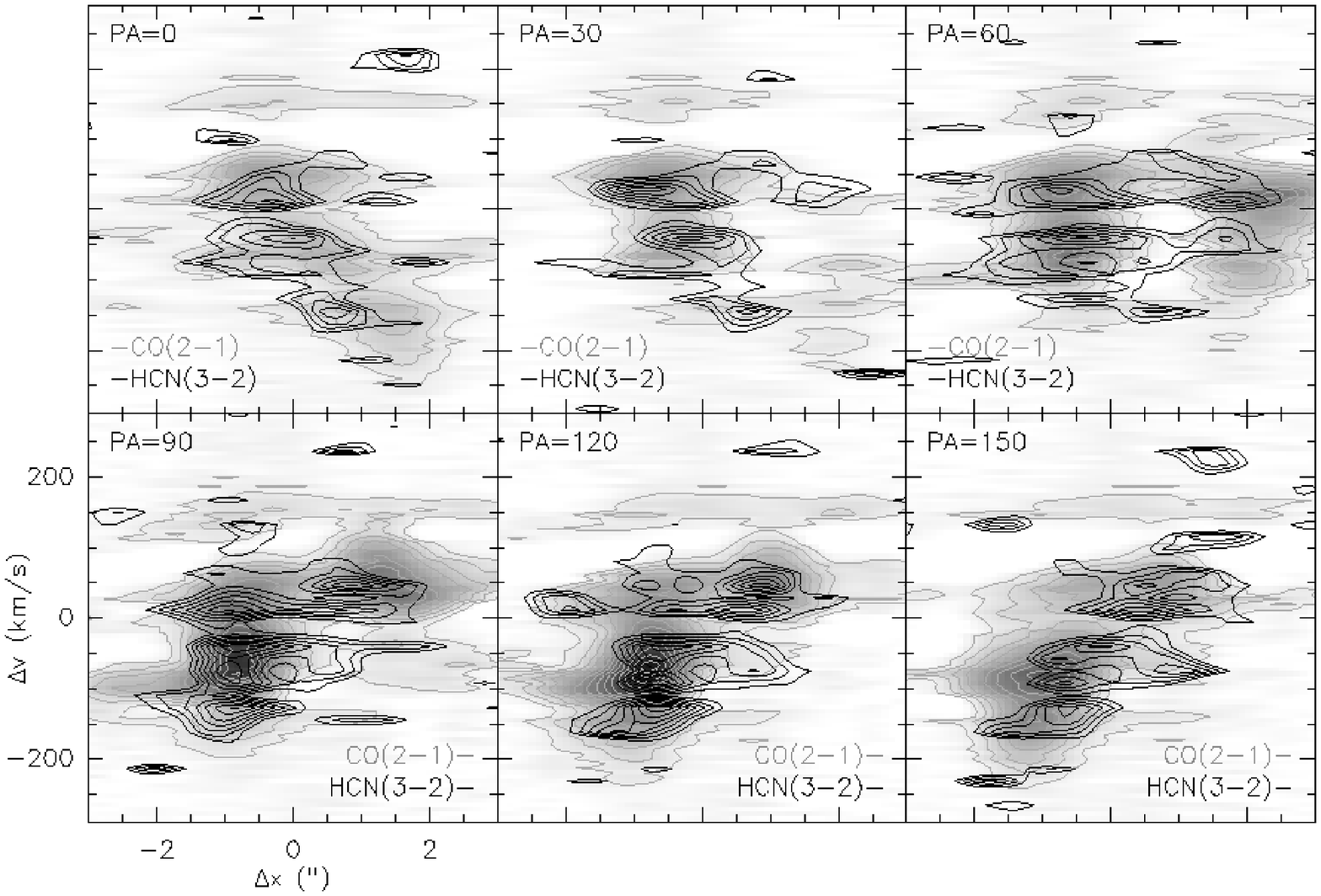}}}
\caption{Position-velocity diagram of NGC~1068 for the
$^{12}$CO(J=2--1) (grey scale), $^{12}$CO(J=3--2) (black contours, top
figure) and HCN(J=3--2) (black contours; bottom figure) emission.}
\label{fig10.0}
\end{figure*}

\begin{figure*}[!t]
\centering
\rotatebox{0}{\resizebox{\hsize}{!}{\includegraphics{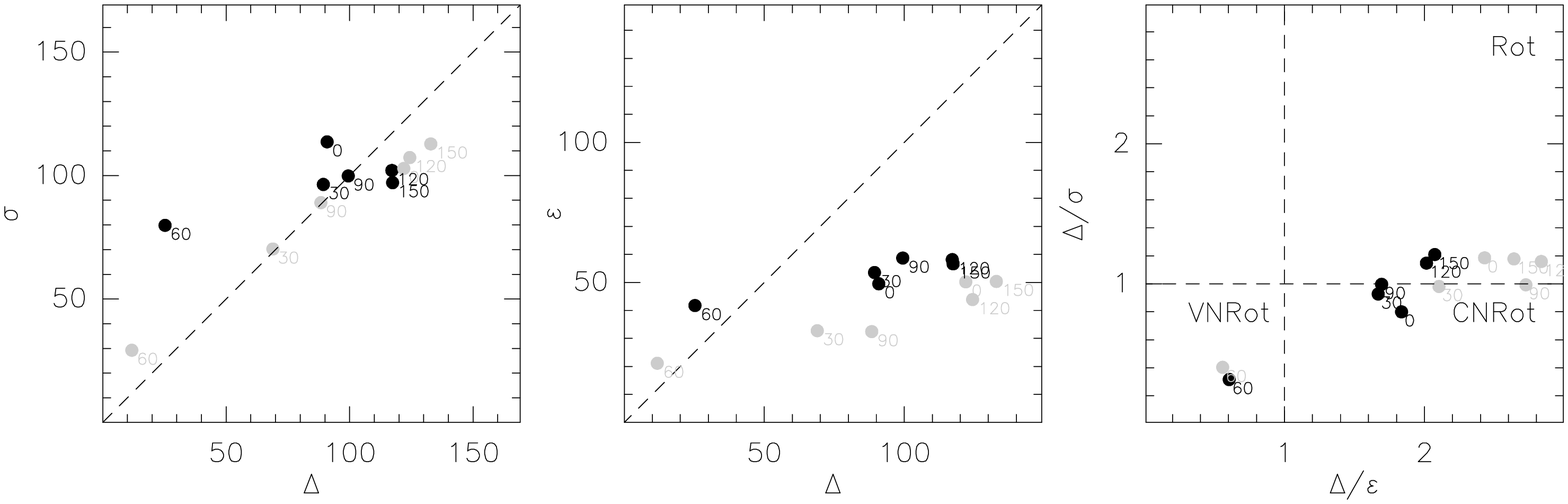}}}
\caption{Parametrisation of the kinematics in the CND of NGC~1068 for
the $^{12}$CO(J=2--1) emission (filled black data points) taken along
the slits at different position angles (at 0, 30, 60, 90, 120, and
150$^\circ$). $\sigma$ is the average line-of-sight velocity
dispersion, $\epsilon$ the rms variation of the velocity from point to
point and $\Delta$ the rotational velocity along each slit (see text
in Section\ref{Dyn} for a more detailed description). VNRot~=~violent
non-rotators, CNRot~=~calm non-rotators and Rot~=~rotators. The filled
grey data points were determined from the model discussed in
Section~\ref{Dyn}.}
\label{fig10.5}
\end{figure*}

The kinematic behaviour of the different molecules is presented in
detail in Fig.~\ref{fig4.0} to \ref{fig10.5}.  To better understand
the puzzling complexity of the different profiles of the various
molecular lines and test whether it is due to dynamical effects, we
spatially split the spectra by deriving the spectrum of the western
and eastern knot separately (Fig.~\ref{fig4.0}). Please note that the
$^{13}$CO(J=1--0) and C$^{18}$O(J=2--1) line emission were discarded
because of their insufficent sensitivity and/or lack of emission in
the western knot while $^{12}$CO(J=1--0) and $^{13}$CO(J=3--2) are not
included because of their insufficient angular resolution. The
iso-velocity maps (Fig.~\ref{fig5.0}) of the $^{12}$CO, $^{13}$CO, HCN
and HCO$^+$ line emission clearly show a dynamical structure that
seems to be dominated by standard disk rotation with a blueshifted
eastern knot and a redshifted western knot. If disk rotation were the
only underlying kinematics, one would expect to find a simple
blueshifted peak at the eastern knot and a redshifted peak at the
western knot. Although disk rotation is observed, Fig.~\ref{fig4.0}
shows kinematic features significantly differing from simple rotation.
Instead, the blueshifted eastern knot also exhibits redshifted
emission and the redshifted western knot blueshifted emission. These
'wings' appear to be present in all three $^{12}$CO emission at a high
significance level as well as in the $^{13}$CO, HCN and HCO$^+$
emission but, given the lower signal-to-noise ratio (SNR) for these
lines, not as pronounced as for $^{12}$CO. At this point, it should be
emphasized that in such a case the moment one map can be very
misleading as it derives only the dominant kinematic structure and
might overlook more complex underlying kinematics. Integrating (in
velocity) the red- and blue-shifted parts of the line spectrum
(Fig.\ref{fig6.0}) as well as analysing the channel maps
(Fig.~\ref{fig7.0} \& \ref{fig8.0}) might be the more appropriate
approach. Fig.~\ref{fig6.0} indicates a more complex distribution than
expected from simple disk rotation.  We find blueshifted emission
spatially coincident with redshifted emission and vice versa; this
seems to be most pronounced for the $^{12}$CO, $^{13}$CO and HCN
emission. By looking at the channel maps of the $^{12}$CO(J=2--1) and
$^{12}$CO(J=3--2) emission (which have the highest SNR), the red- on
blueshifted and blue- on redshifted emission is not only at
low-velocities but also at higher velocities (which is especially
visible in the $^{12}$CO(J=2--1) emission; see channels $<$$-$70km/s
and $>$+100~km/s in Fig.~\ref{fig7.0}). The same is true for the
$^{12}$CO(J=3--2) emission (Fig.~\ref{fig8.0}, although less
pronounced, especially for velocities $>$80~km~s$^{-1}$ for which no
emission can be found anymore as opposed to $^{12}$CO(J=2--1)). We
find a behaviour of the HCN(J=3--2) emission similar to that of the
$^{12}$CO(J=2--1) and $^{12}$CO(J=3--2) emission though on a much
lower significance level. The $^{13}$CO(J=2--1) emission seems to
indicate, however, a different behaviour (see
Fig.~\ref{fig9.0}). Instead of being distributed in a 'ring'-like
manner, the emission appears to be elongated more in a South-West to
North-East direction (see especially channels maps between
$+$50~~km~s$^{-1}$ to $-$20~km~s$^{-1}$). However, given the low
sensitivity level, this structure has to be treated with caution and
needs confirmation by either higher sensitivity observations or other
molecular lines such as SiO. Indeed, the SiO emission seems to
indicate a similar behaviour as discussed in separate papers
\citep[see][]{garc08,garc10}.

The position-velocity diagrams of the $^{12}$CO and HCN emission,
taken at different Position Angles (PA) in steps of 30$^\circ$ across
the CND, are shown in Fig.~\ref{fig10.0}. The grey scale denotes the
$^{12}$CO(J=2--1) emission for better comparison. Overall, the
kinematic structures in the different lines strongly resemble each
other. Also, the overlap of the red- on blueshifted emission can be
seen quite well in the position-velocity diagrams (see especially
PA=60-120$^\circ$ in Fig.~\ref{fig10.0}), strongly indicating
pronounced non-circular motions in the CND of NGC~1068.

In order to quantify and parametrise the observed complex kinematics,
we follow the approach used by \cite{heck89} and \cite{baum92}.  We
determine three kinematic parameters from the slits taken at the
different position angles used in Fig.~\ref{fig10.0} for the $^{12}$CO
emission. These parameters are: 1.) the average line-of-sight velocity
dispersion $\sigma$, determined as 0.426$\times$FWHM along each slit,
2.) the ``rotational'' velocities $\Delta$, determined from the
difference between the average velocities on either side of the
nucleus along each slit, and 3.)  the rms variation of the velocity
$\epsilon$ for each point along the slit, defined as
$\epsilon=\sqrt{\frac{1}{N}\Sigma_{i=1}^{N}(v_i-v_{\rm av})^2}$, where
$N$ is the number of points along the slit, $v_i$ the intensity
weighted velocity for point $i$ and the $v_{\rm av}$ the intensity
weighted average velocity along the slit. A comparison of the
different parameters with each others, especially the ratios
$\frac{\Delta}{\sigma}$ and $\frac{\Delta}{\epsilon}$ allows to
classify the dynamics into three different groups (for a more detailed
explanation we refer to \cite{baum92}):

\begin{itemize}
\item[] ROTATORS: $\Delta/\sigma\gtrsim1$,
                  $\Delta/\epsilon\gtrsim1$
\item[] CALM NON-ROTATORS: $\Delta/\sigma<1$,
                           $\Delta/\epsilon\sim1$
\item[] VIOLENT NON-ROTATORS: $\Delta/\sigma<1$,
                              $\Delta/\epsilon<1$
\end{itemize}

The results of this kinematical parametrisation for the
$^{12}$CO(J=2--1) are shown in Fig.~\ref{fig10.5}, highlighting our
previous findings (we find very similar values when taking the
$^{12}$CO(J=3--2) emission). For most position angles, we find that
the ratios are most consistent with calm non-rotation with one
exception (PA=60$^\circ$) being located in the area of the violent
non-rotators. This strongly emphasizes the fact that although there is
an underlying dominating disk rotation, the dynamics of the molecular
gas in the CND of NGC~1068 is significantly disturbed by a
non-rotational process, most significantly for the North,
North-Eastern part of the CND (i.e., along PA=0-90$^\circ$).  We also
attempted to determine the rotational velocities directly from
azimuthally averaging the velocities within the CND and then
subsequently fitting rotational curves to the data using rotcurv in
gipsy. We thereby assumed different scenarios ranging from pure disk
rotation to adding a radial dependency. However, no physically
meaningful result could be obtained. We mostly find positive
velocities that decrease with radius which is inconsistent with simple
disk rotation \cite[see also the dynamical analysis done
in][]{schin00} and necessitates the inclusion of a bulge or central
mass component whether in form of nuclear star cluster or a massive
black hole in addition to the disk. This certainly further emphasizes
the complexity of the molecular gas dynamics in the CND of NGC~1068.

\subsubsection{Molecular Line Ratios}
\begin{figure*}[!]
\centering
\rotatebox{0}{\resizebox{\hsize}{!}{\includegraphics{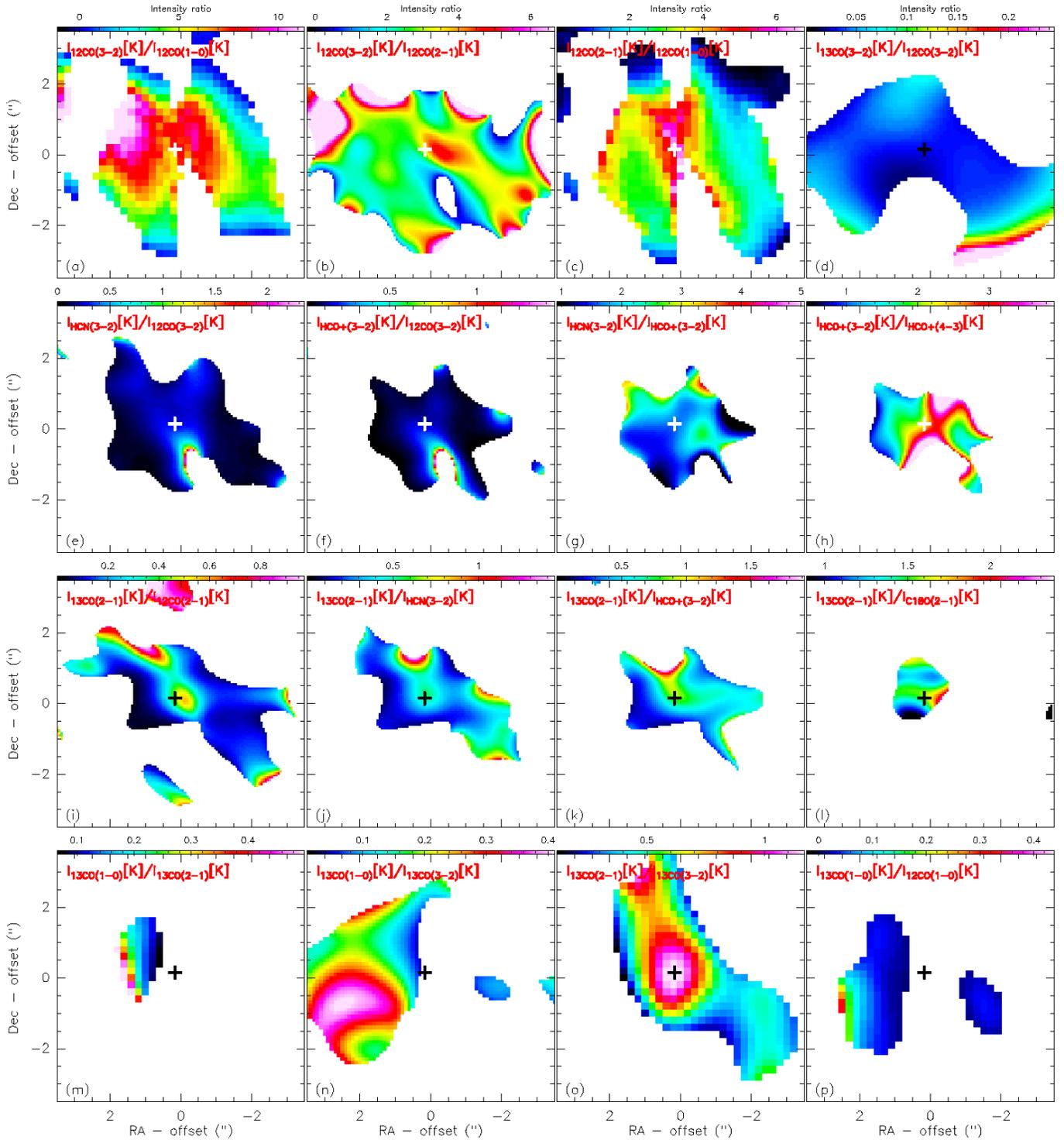}}}
\caption{Velocity integrated molecular line intensity ratios for
  NGC~1068 above a $\geq$1$\sigma$ threshold. The white cross marks
  the position of the mm continuum emission that is associated with
  the AGN. The different line ratios are: {\it a)}
  $^{12}$CO(J=3--2)-to-$^{12}$CO(J=1--0); {\it b)}
  $^{12}$CO(J=3--2)-to-$^{12}$CO(J=2--1), {\it c)}
  $^{12}$CO(J=2--1)-to-$^{12}$CO(J=1--0), {\it d)}
  $^{12}$CO(J=3--2)-to-$^{13}$CO(J=3--2), {\it e)}
  HCN(J=3--2)-to-$^{12}$CO(J=3--2), {\it f)}
  HCO$^+$(J=3--2)-to-$^{12}$CO(J=3--2), {\it g)}
  HCN(J=3--2)-to-HCO$^+$(J=3--2), {\it h)}
  HCO$^+$(J=3--2)-to-HCO$^+$(J=4--3), {\it i)}
  $^{13}$CO(J=2--1)-to-$^{12}$CO(J=2--1), {\it j)}
  $^{13}$CO(J=2--1)-to-HCN(J=3--2), {\it k)}
  $^{13}$CO(J=2--1)-to-HCO$^+$(J=3--2), {\it l)}
  $^{13}$CO(J=2--1)-to-C$^{18}$O(J=2--1), {\it m)}
  $^{13}$CO(J=1--0)-to-$^{13}$CO(J=2--1), {\it n)}
  $^{13}$CO(J=1--0)-to-$^{13}$CO(J=1--0), {\it o)}
  $^{13}$CO(J=2--1)-to-$^{13}$CO(J=3--2), {\it p)}
  $^{13}$CO(J=1--0)-to-$^{12}$CO(J=1--0)}
\label{fig11.0}
\end{figure*}

\begin{figure*}[!t]
\centering
\rotatebox{-90}{\resizebox{!}{\hsize}{\includegraphics{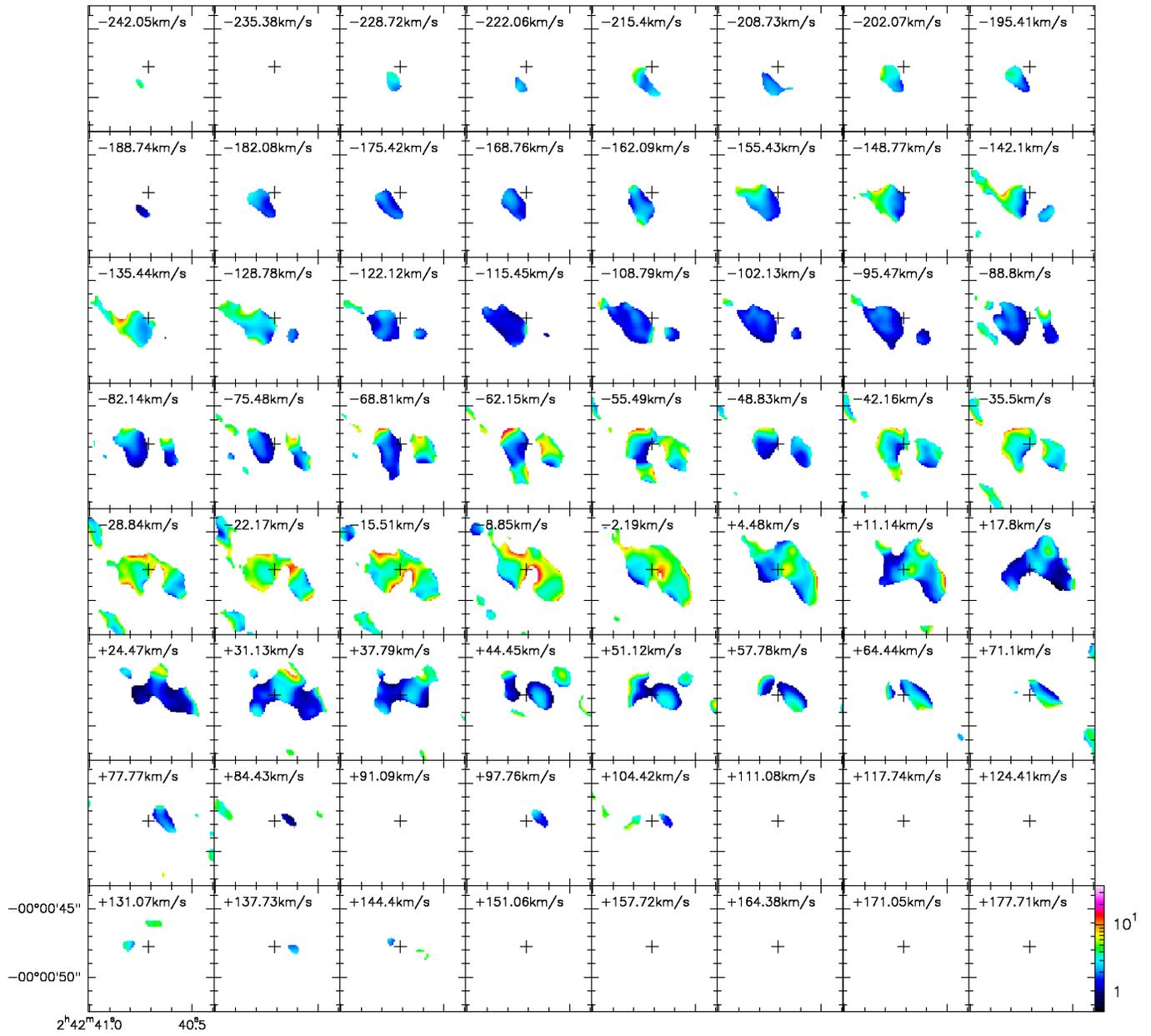}}}
\caption{$^{12}$CO(J=3--2)-to-$^{12}$CO(J=2--1) line ratio above
  a 4$\sigma$-threshold for each line. }
\label{fig12.0}
\end{figure*}

\begin{figure*}[!t]
\centering
\rotatebox{-90}{\resizebox{!}{\hsize}{\includegraphics{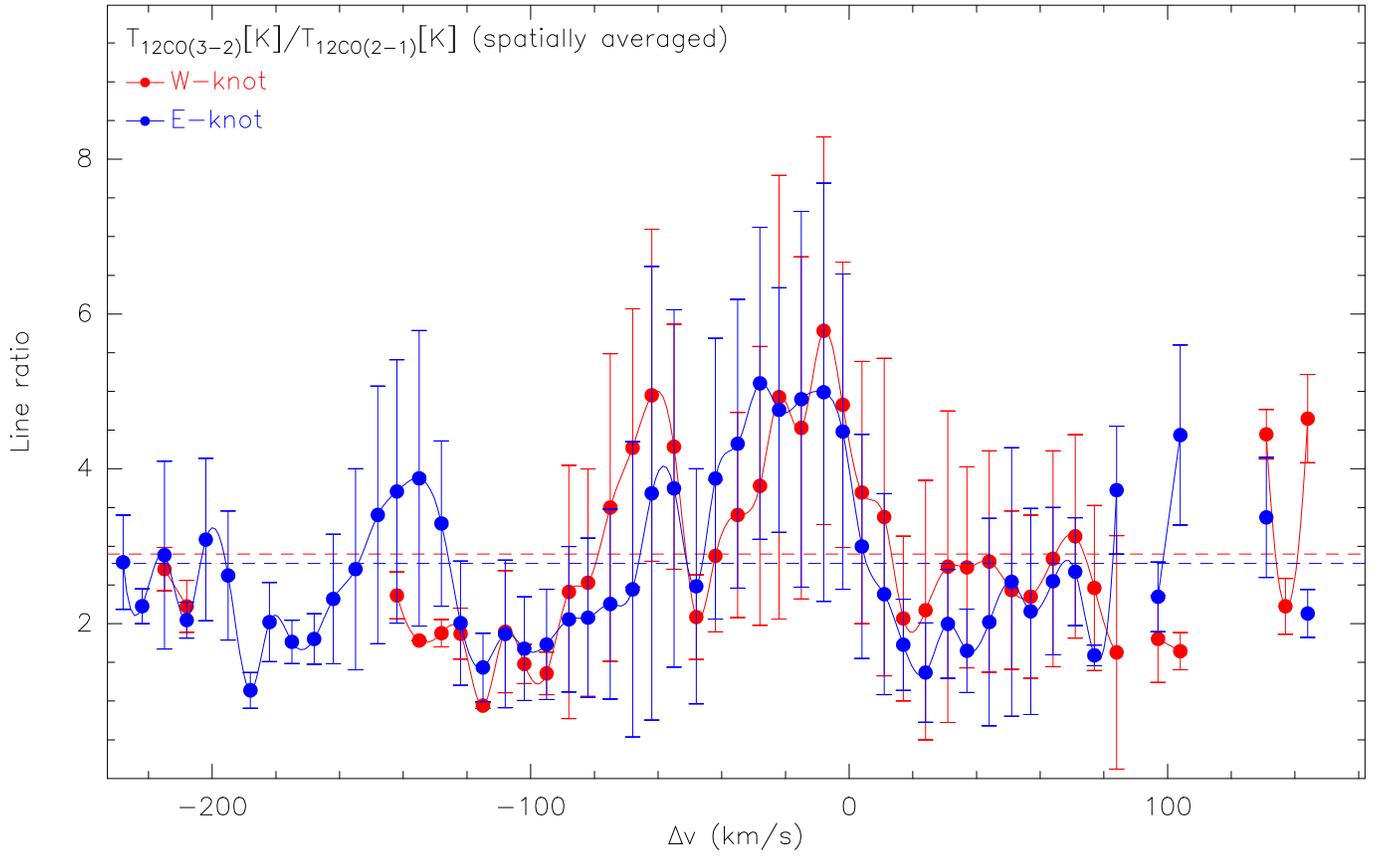}}}
\caption{Spatially averaged line ratio for the eastern and western
knot as function of velocity, derived from Fig.~\ref{fig12.0}. The
error bars denote the variance of each averaged value, which does not
exceed 50\% in most cases. The missing values for velocities $-$120
and $-$200~km~s$^{-1}$ and $+$100 to $+$140~km~s$^{-1}$ are due to the
lack of emisison in the respective knot. The dashed lines represent
the median of the line ratios for the two knots.}
\label{fig13.0}
\end{figure*}

In order to constrain the excitation conditions and chemistry of the
molecular gas, we derive the line ratios for the different molecules
and transitions in several ways, by accounting for the different
angular resolutions (especially with respect to the $^{12}$CO(J=1--0),
$^{13}$CO(J=1--0) and $^{13}$CO(J=3--2) emission). Before determining
any line ratio, the line emission was brought to the same (lower)
angular resolution by using a uv-taper; this seems appropriate for
most of the lines as we recovered most of the emission with the
interferometric observations. Also, we compute line ratios only for
emission coming from the samwe spatial regions (see
Figs.\ref{fig11.0}-\ref{fig13.0}). Fig.~\ref{fig11.0} shows the
velocity integrated line ratios between various combinations of the
molecular lines.  Separating spatially the eastern and western knot we
derive spatially averaged line ratios from Fig.~\ref{fig11.0} which
are listed in Table~\ref{tab6}.  Please note that the values in
Table~\ref{tab6} might vary from those derived from Tables~\ref{tab4}
and \ref{tab5}. However, the differences can be easily explained by
the different sequence of averaging (i.e., first in space, then in
velocity versus first in velocity, then in space) for
Table~\ref{tab5}, the usage of an elliptical Gaussian
fit\footnote{Although this a reasonable first order fit, the line
emission is certainly not exactly of an elliptical Gaussian shape so
that some of the emission is not well reproduced by fitting an
elliptical Gaussian to the velocity integrated maps.} for
Table~\ref{tab4} as opposed to spatially averaging without a fit as
done for Table~\ref{tab6}, and the different spatial resolutions of
the line emission in Tables~\ref{tab4} \& \ref{tab5}.

We identify some spatial variance of the different line ratios (mostly
a factor $\sim$2-3) which seem to be most pronounced in the $^{12}$CO
and $^{13}$CO line ratios (Fig.~\ref{fig11.0}a,d,o). The HCN and
HCO$^+$ line ratios seem to be more constant over the CND than
$^{12}$CO and $^{13}$CO. The higher values are found closer to the
position of the AGN for most maps.

In order to investigate whether there might be a velocity (and
spatial) dependence on the line ratios, we determined the line ratio
channel maps for the two strongest $^{12}$CO transitions (J=2--1 and
J=3--2) as function of velocity in Fig.~\ref{fig12.0}. We find
somewhat higher (i.e., factor of $\gtrsim$2)
$^{12}$CO(J=3--2)-to-$^{12}$CO(J=2--1) ratios close to the AGN at
velocities around the systemic velocity but also on the eastern knot
at high negative velocities ($\lesssim$$-$130~km~s$^{-1}$). Both knots
seem to show (more or less) the same velocity and spatial behaviour as
can be seen in Fig.~\ref{fig13.0}. This plot shows the spatially
averaged $^{12}$CO(J=3--2)-to-$^{12}$CO(J=2--1) line ratios for the
eastern and western knot as function of velocity. The two curves
follow each other nicely except for velocities between $-$120 to
$-$140~km~s$^{-1}$ for which the eastern knot exposes higher values
(by a factor of 2). The error bars denote the variance of each
averaged value which in most cases indicates a variation by a factor
of 1.5.

\section{Discussion}
\label{dis}

\subsection{Spectral Energy Distribution of the Continuum Emission}
\label{sed}
The nature of the continuum emission (from IR over sub-mm to cm
wavelengths) represents a highly debated and complicated matter for
NGC~1068, recently gaining a revival by newly published VLTI/MIDI
(i.e., IR) and radio data \citep[e.g.,][]{hoen08,cott08}. As mentioned
in the previous section, the radio continuum emission splits up into
several components, a jet plus counter-jet and a core component (S1)
associated with the AGN itself. While the emission from the jet is
certainly pure non-thermal synchrotron emission, as supported by its
steep continuum spectrum \citep[e.g.,][]{gall04,cott08}, the nature of
the emission from S1 is highly controversial. \cite{gall04} already
rule out synchrotron emission as origin for the continuum spectrum of
S1 and discuss electron-scattered synchrotron emission as well as
thermal free-free absorption as alternatives. While \cite{krip06}
present arguments for electron-scattered synchrotron emission based on
a turnover seen between cm and mm-data, \cite{cott08} rather support
the thermal free-free absorption model.  A highly complicating factor
in this discussion is certainly the mismatch in angular resolution
between the cm, mm and IR data.  As discussed in \cite{krip06} and
this paper, the mm-continuum emission is contaminated by emission from
the jet at angular resolutions of $\gtrsim$1$''$, introducing large
uncertainties in the estimate of S1's flux \citep[compare][]{hoen08,
cott08,krip06} due to the lack of angular resolution. However, the
previous estimate of the 1.3~mm continuum flux of (10$\pm$4)~mJy in
\cite{krip06}, translating to (9$\pm$4)~mJy at 1.0~mm, is very similar
to the {\it measured} 1.0~mm continuum flux of (13$\pm$2)~mJy from our
high-angular resolution SMA observations. Given the unresolved nature
of the latter, jet emission does not seem to be significant anymore at
this angular resolution \citep[see also Fig.~1 in][]{cott08} although
the obtained angular resolution is still an order of magnitude larger
than that at cm wavelengths.

\begin{figure}[!]
\centering
\rotatebox{0}{\resizebox{9cm}{!}{\includegraphics{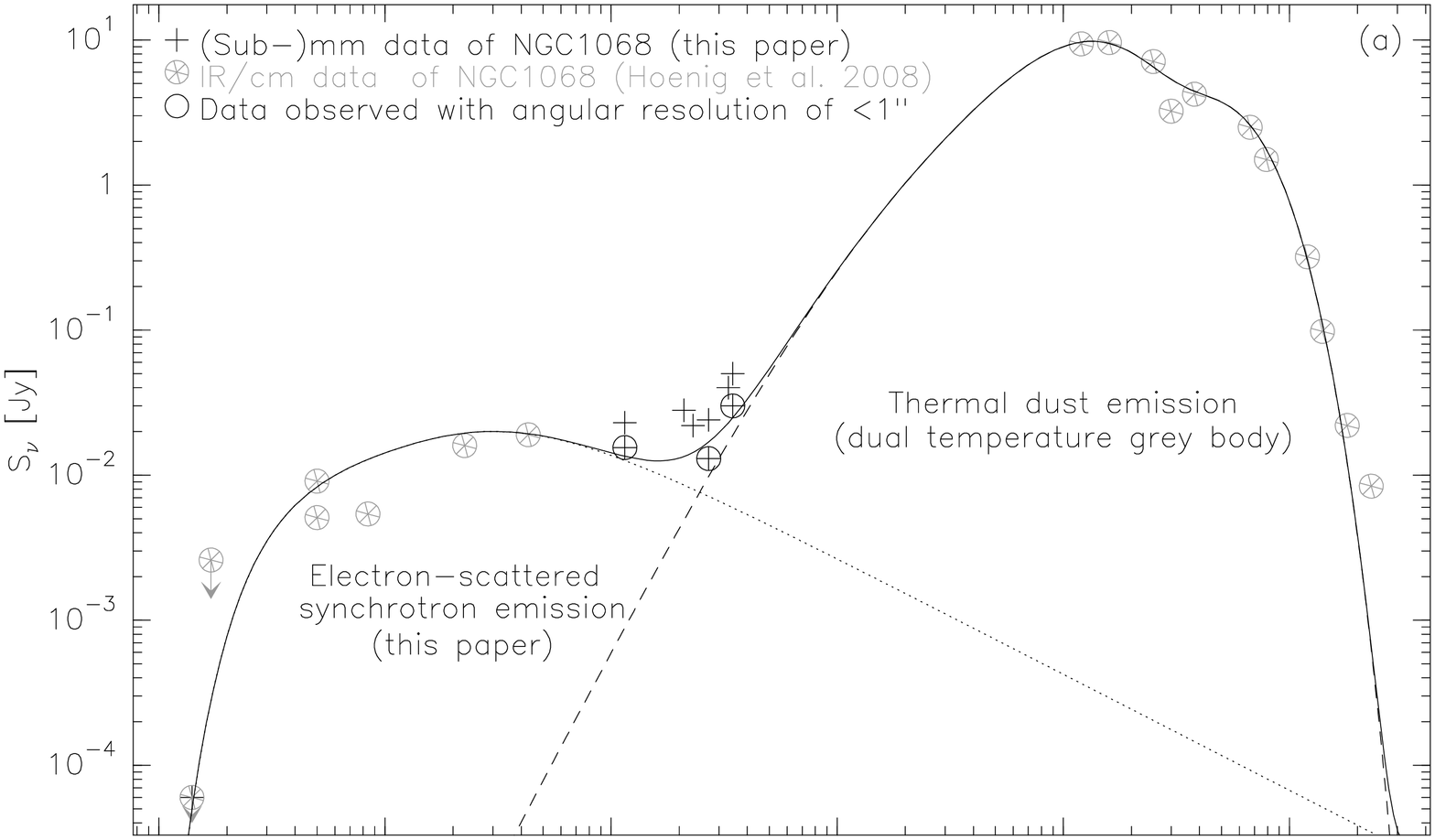}}}\\
\rotatebox{0}{\resizebox{9cm}{!}{\includegraphics{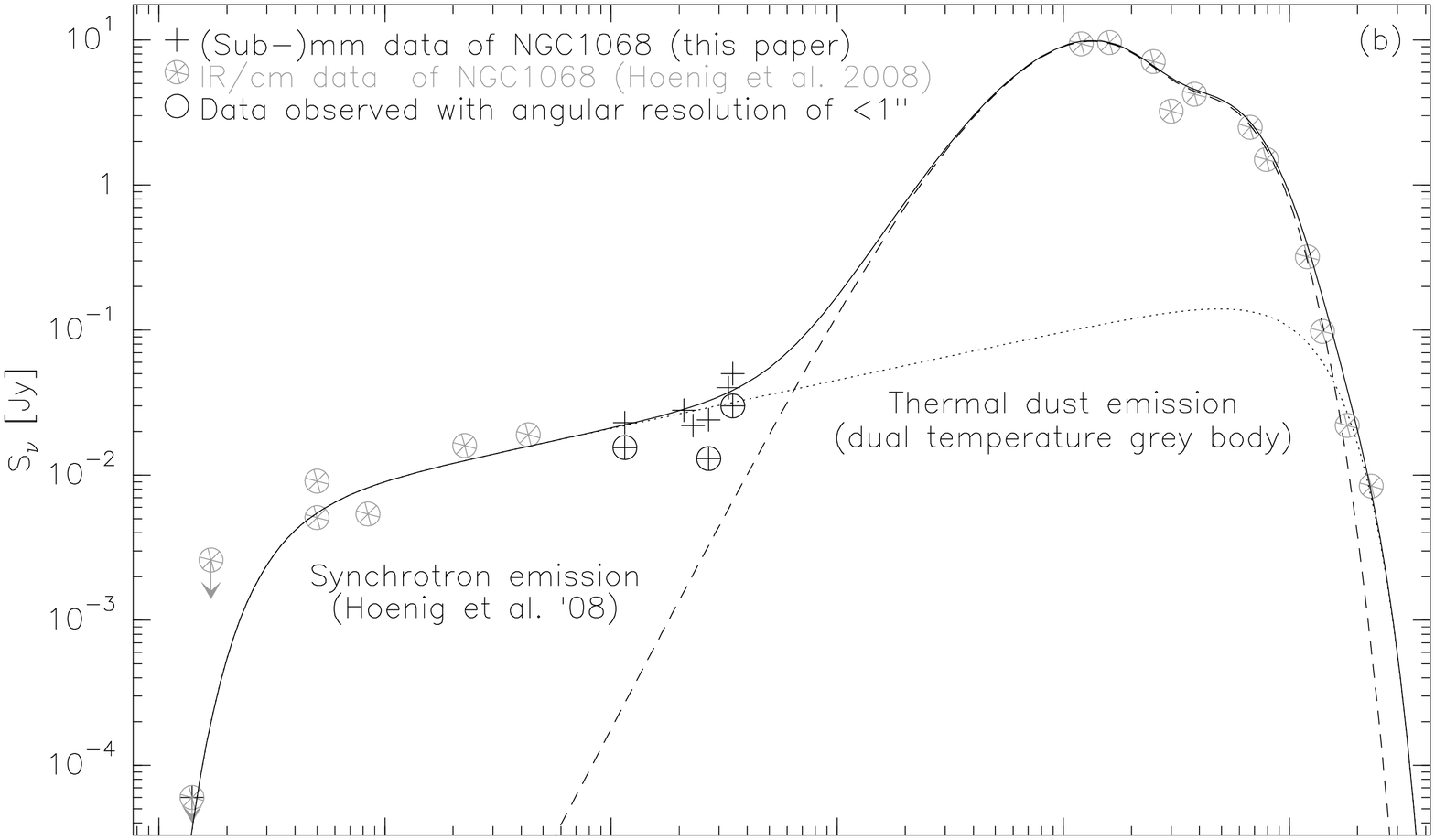}}}\\
\rotatebox{0}{\resizebox{9cm}{!}{\includegraphics{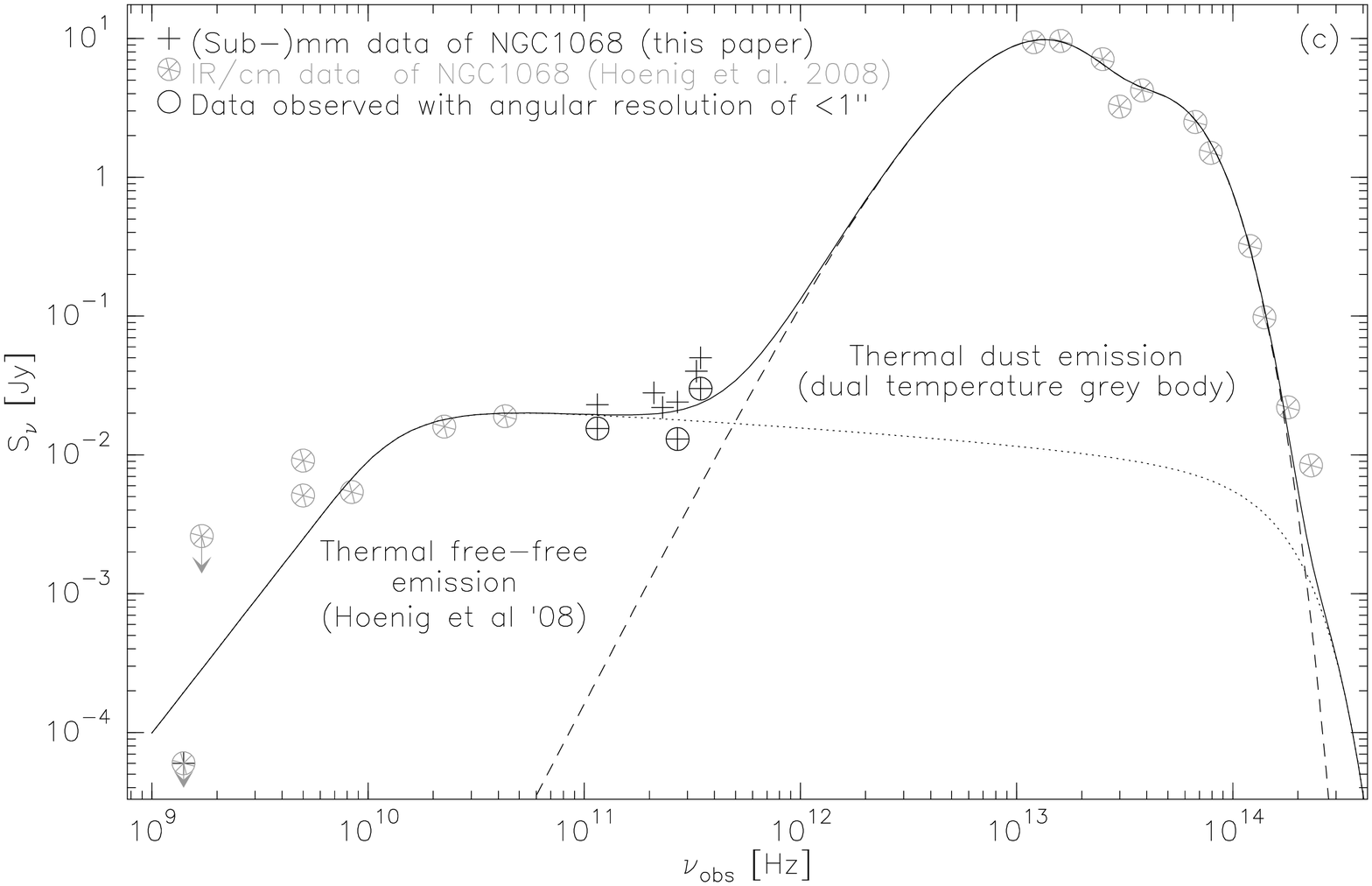}}}
\caption{Spectral energy distribution of the continuum emission in
NGC~1068, based on data from this paper, \cite{krip06} ({\it black
crosses}), and \cite{hoen08} ({\it grey crosses}). The dotted line
respresents the model for the radio continuum emission (either
electron scattered synchrotron emission ({\it a}), synchrotron
emission ({\it b}), and free-free absorption ({\it c})), the dashed
line represents the model for the IR data (a two-temperature grey body
{\it a-c}), and the solid line represents the composite of both ({\it
a-c}). The data observed at an angular resolution below 1$''$ are
additionally marked with a circle.}
\label{fig14.0}
\end{figure}

Taking also the new 850~$\mu$m (UN) continuum flux measurements into
account, we recomputed the spectral energy distribution (SED) and
replotted the models used by \cite{hoen08} and \cite{krip06}. We
thereby base our graphs on the formula and parameters specified in
Equations (2)-(5) and Table~2\&3 in \cite{hoen08} and Equations
(1)-(2) and Fig.~3 in \cite{krip06}. The results are shown in
Fig.~\ref{fig14.0}a-c. We marked all data points with a circle for
which the obtained angular resolution of the observations did not
exceed 1$''$\footnote{Please note, that this is true for all data
points except that at 3~mm. The encircled S1 data point has been
estimated at 3~mm, not observed. We added this data point for
consistency reasons only.}.  We also fitted a two-temperature grey
body to the IR data, in order to estimate the contribution of thermal
dust emission to the sub-mm continuum emission. Although this
grey-body fit is certainly not as sophisticated as the clumpy torus
model used in \cite{hoen08}, it represents a reasonable approximation
as demonstrated by the good match to the IR data points.

As can be seen in Fig.~\ref{fig14.0}b and c, the models used by
\cite{hoen08} significantly overestimate the observed 1~mm (UN) flux
by a factor of 2-3, although they reproduce correctly the 850~$\mu$m
one. It seems that the electron-scattered synchrotron emission model
in Fig.~\ref{fig14.0}a and the thermal free-free emission model in
Fig.~\ref{fig14.0}c both need the extra contribution from the thermal
dust emission to correctly reproduce the 850~$\mu$m (UN) flux, while
the synchrotron model in Fig.~\ref{fig14.0}b does not require it.
Thus, it appears to be indeed very likely that the continuum emission
is dominated by thermal dust emission starting at wavelengths
$\lesssim$850$\mu$m, as posited above.

Based on the 1~mm (UN) flux, it seems that the model best reproducing
the SED at cm and mm wavelengths is the electron-scattered synchrotron
emission; the thermal free-free absorption seems to overpredict the
1~mm flux. However, observations of the continuum emission in NGC~1068
have to be conducted at similarly high angular resolutions
($\ll$0.5$''$), in order to dispel all remaining doubts, although the
new mm observations presented in this paper are already a step in the
right direction.

\subsection{Dynamics of the Molecular Gas}
\label{Dyn}

In previous studies, the complex kinematic behaviour of the molecular
gas has been thought to be a consequence of a warped disk. The warped
disk has been modelled with a tilted ring model
\citep[e.g.,][]{schin00}. However, the spatial overlap between the
red- and blueshifted emission (i.e., the existence of highly
non-circular motions) cannot be reproduced by these tilted ring models
because they are based on circular motions and thus cannot account for
non-circular motions of the gas (within the plane).

Even though we cannot rule out a warped disk scenario in which part of
the gas could be trapped in elliptical orbits producing the
non-circular motions, we want to propose an alternative approach,
following recent findings on the H$_2$(1--0) S(1) emission at scales
of 100-150pc by \cite{mueller09} and the model proposed by
\cite{galli02}. The nature of the dynamics displayed in
Fig.~\ref{fig4.0}, \ref{fig5.0}, \ref{fig6.0}, \ref{fig7.0} and
\ref{fig8.0} could also be explained by the following scenario: a
rotating disk plus an outflow of the disk gas due to shocks and/or a
CND-jet interaction.  This hypothesis seems to further gain support
when considering besides the H$_2$ 1--0 S(1) map \citep[][see their
Fig.~4]{mueller09}, also the 12$\mu$m map \citep[][see their
Fig.~4]{bock00} and the 5cm radio-continuum map \citep[][see their
Fig.~1]{gall04}. The H$_2$ 1--0 S(1) and 12$\mu$m emission follow
nicely that of the radio jet in the inner 1$''$ (North/North-East
direction) which seems to interact with emission from the molecuar gas
in the CND at $\pm$1-2$''$ in the Northern part (see next
Sub-Section); both the blue and redshifted components associated with
the non-circular motions to the East and West of where the jet goes
through or lies in front of the CND \citep[see also the case of
M51;][]{mats07,mats04}. As the jet shows a biconical structure with a
change in direction close to the CND, it is not unreasonable to
believe that part of it indeed hits the CND \citep[see
also][]{kraem98}. Such an interaction could easily produce a shock
through/along the CND (at least in the northern part, i.e. the bridge
between the eastern and western knot) feeding the assumption that some
part of the molecular gas in the ring/disk might be blown outwards.

Other causes for expanding/shocked gas include hypernovae explosions,
stellar winds from a super stellar cluster as suspected in some nearby
starburst galaxies \citep[such as NGC~253 or M82,][]{saka06,mats00} or
cloud-cloud collision within the CND \citep[i.e., within the inner
Lindblad resonance; see][]{garc10}. However, they seem rather unlikely
since the CND of NGC~1068 does not show any signs of starburst
activity and the expansion seems to be too ``ordered'', i.e., too
symmetric, to be caused by highly chaotic cloud-cloud collisions.
Also, we cannot entirely rule out that the dynamics we see in the
molecular gas is due to inflowing rather than outflowing gas,
especially since indications of inflowing gas along the jet within the
central $\sim$70~pc (i.e., on scales smaller than the CND) have been
already presented in previous studies
\citep[see][]{mueller09}. However, based on the appearance of the
molecular gas within the disk (ring-like structure with an apparent
void of gas in the inner part) lets us favor the jet-gas interaction
rather than an inflow scenario (on scales of 100-150~pc). An outflow
on scales of 100-150~pc is not necessarily in contradiction with an
inflow scenario on scales $\lesssim$70~pc proposed by
\cite{mueller09}. The jet-gas interaction could equally drag gas
outwards on scales larger than 100~pc but trigger an inflow at smaller
scales, depending on the type of interaction between the jet and the
molecular clouds. However, higher-angular resolution observations at
sub-arcsecond angular resolution (as possible with ALMA) will
certainly help to distinguish between the different scenarios.

\begin{figure*}[!t]
\centering
\resizebox{\hsize}{!}{\includegraphics{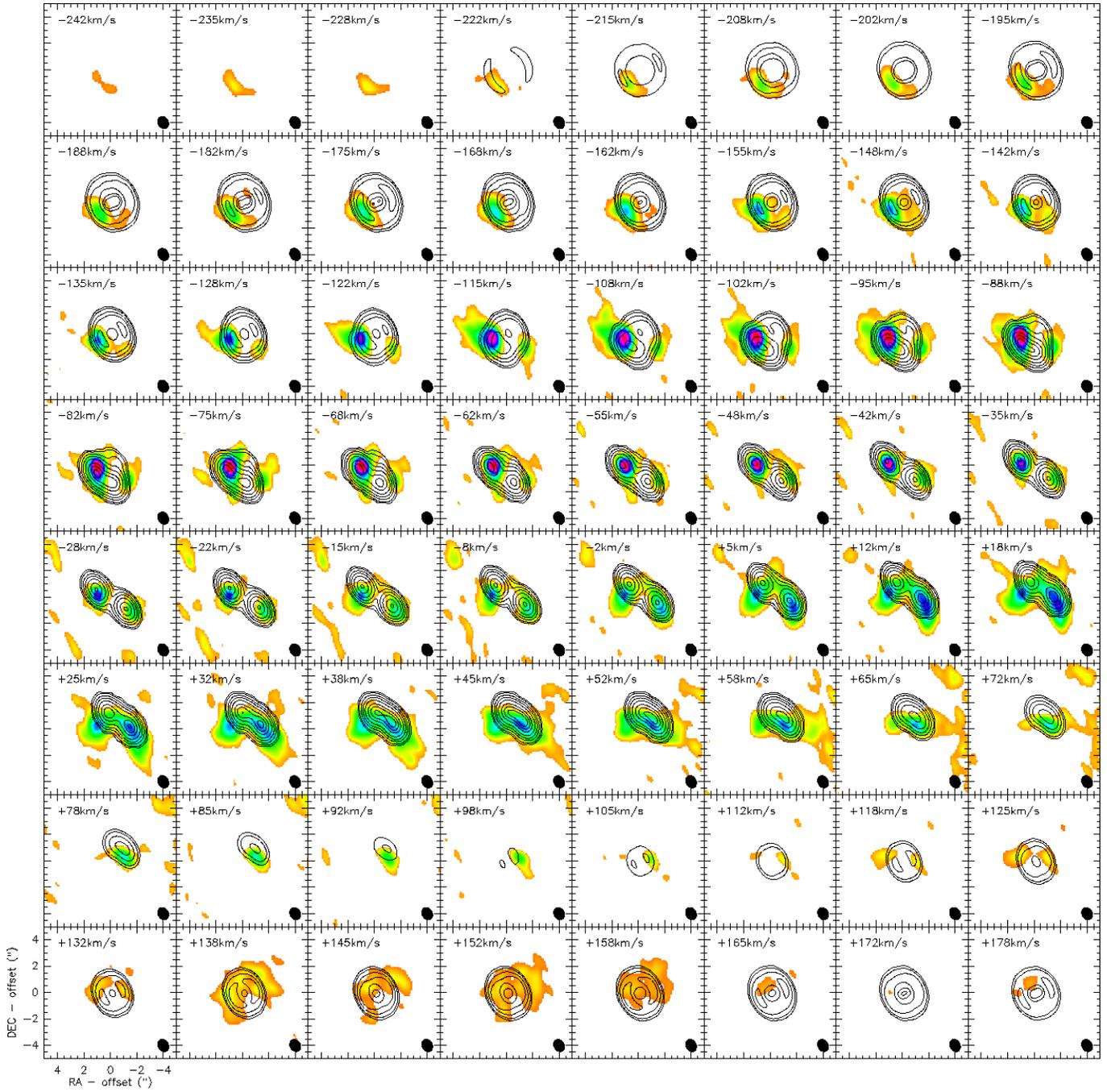}}
\caption{Velocity channel maps of the CO model compared to the
$^{12}$CO(2--1) emission.}
\label{fig15.0}
\end{figure*}

\begin{figure*}[!t]
\centering
\resizebox{12cm}{!}{\includegraphics{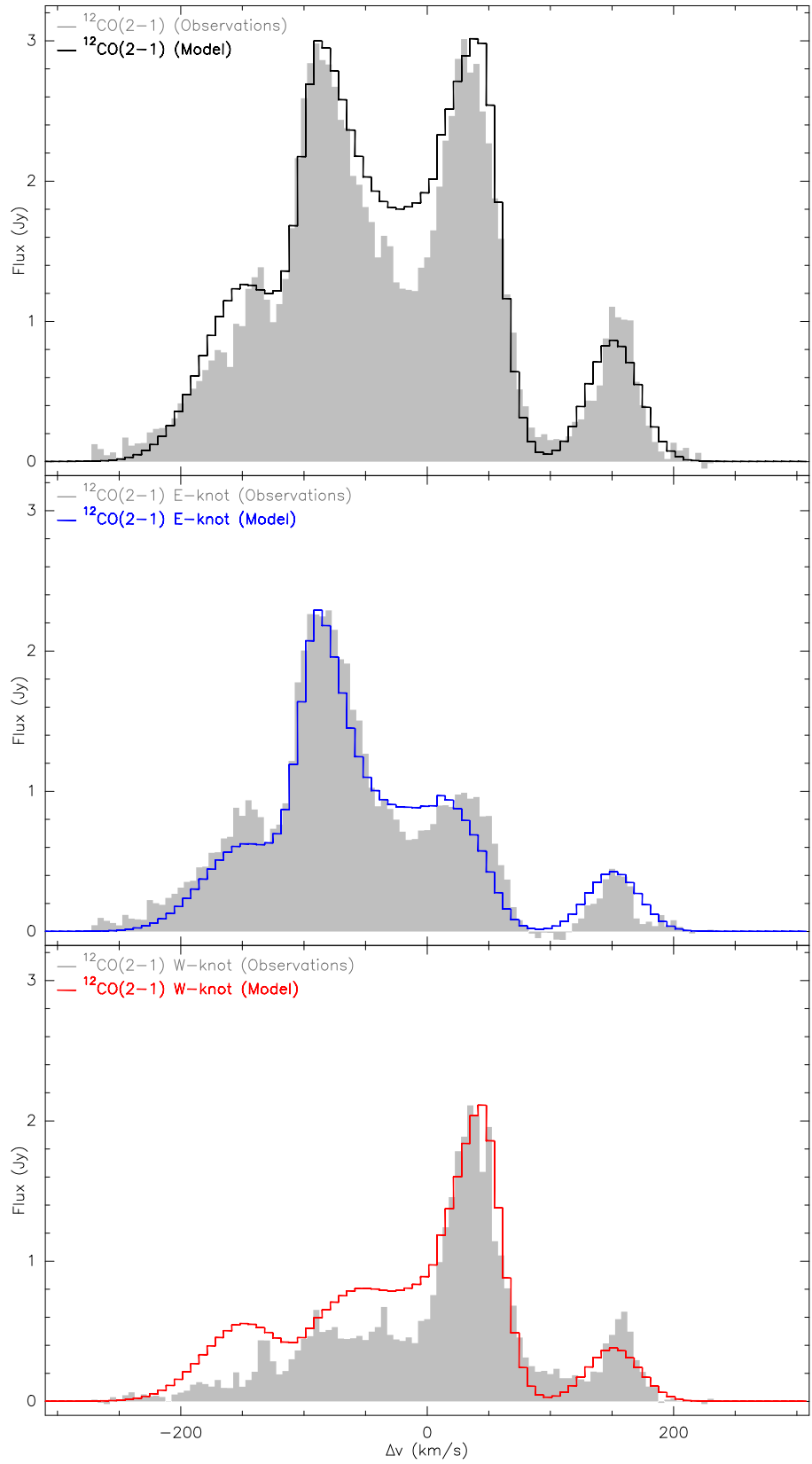}}
\caption{Spectrum of the CO model compared to the $^{12}$CO(2--1)
emission.}
\label{fig16.0}
\end{figure*}

\begin{figure*}[!t]
\centering
\resizebox{15cm}{!}{\includegraphics{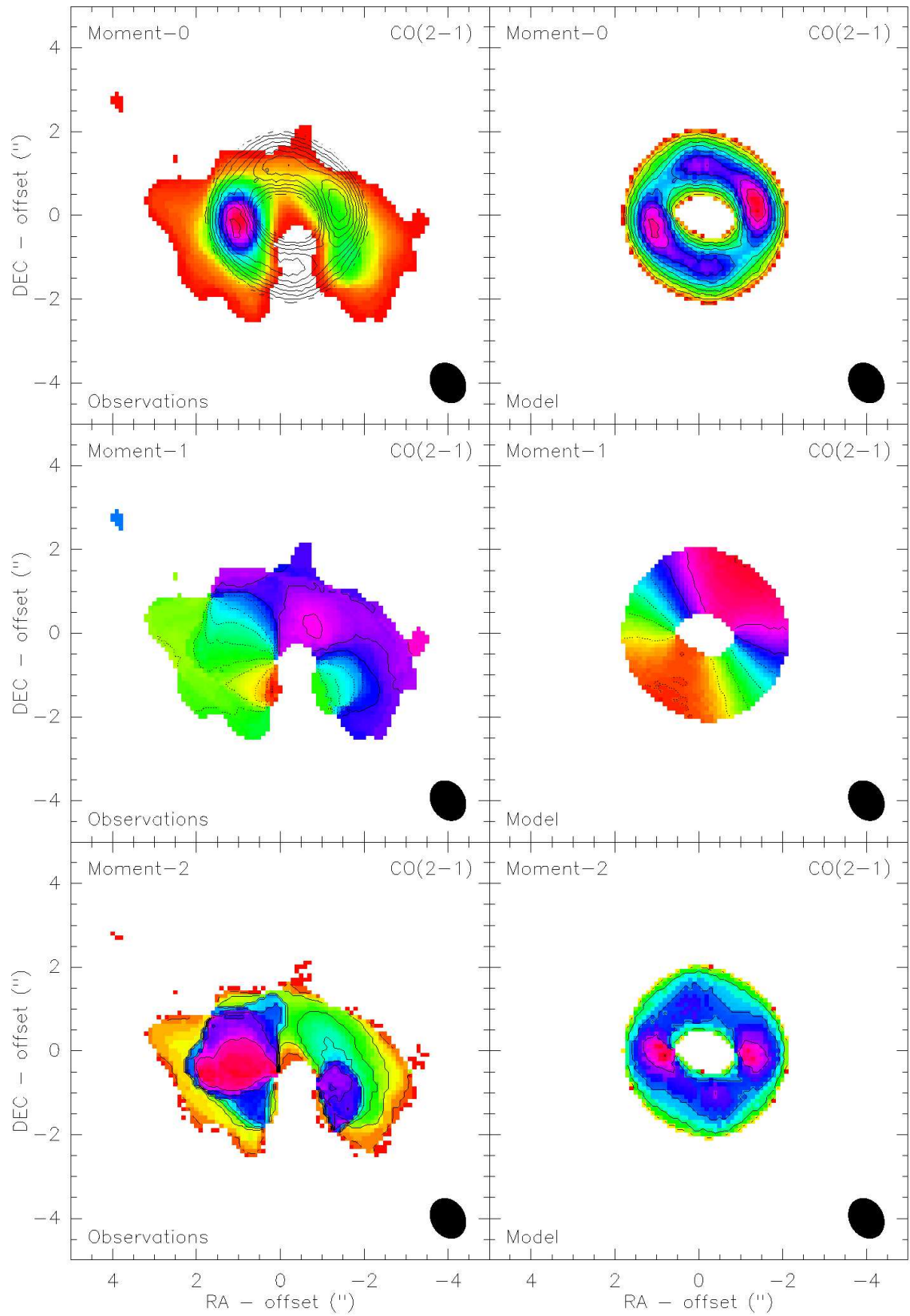}}
\caption{Moment maps of the CO model compared to the $^{12}$CO(2--1)
emission. The velocities are plotted in steps of 20$km~s^{-1}$ for the
Moment-1 and Moment-2 maps.}
\label{fig17.0}
\end{figure*}

\begin{figure*}[!t]
\centering
\rotatebox{-90}{\resizebox{!}{\hsize}{\includegraphics{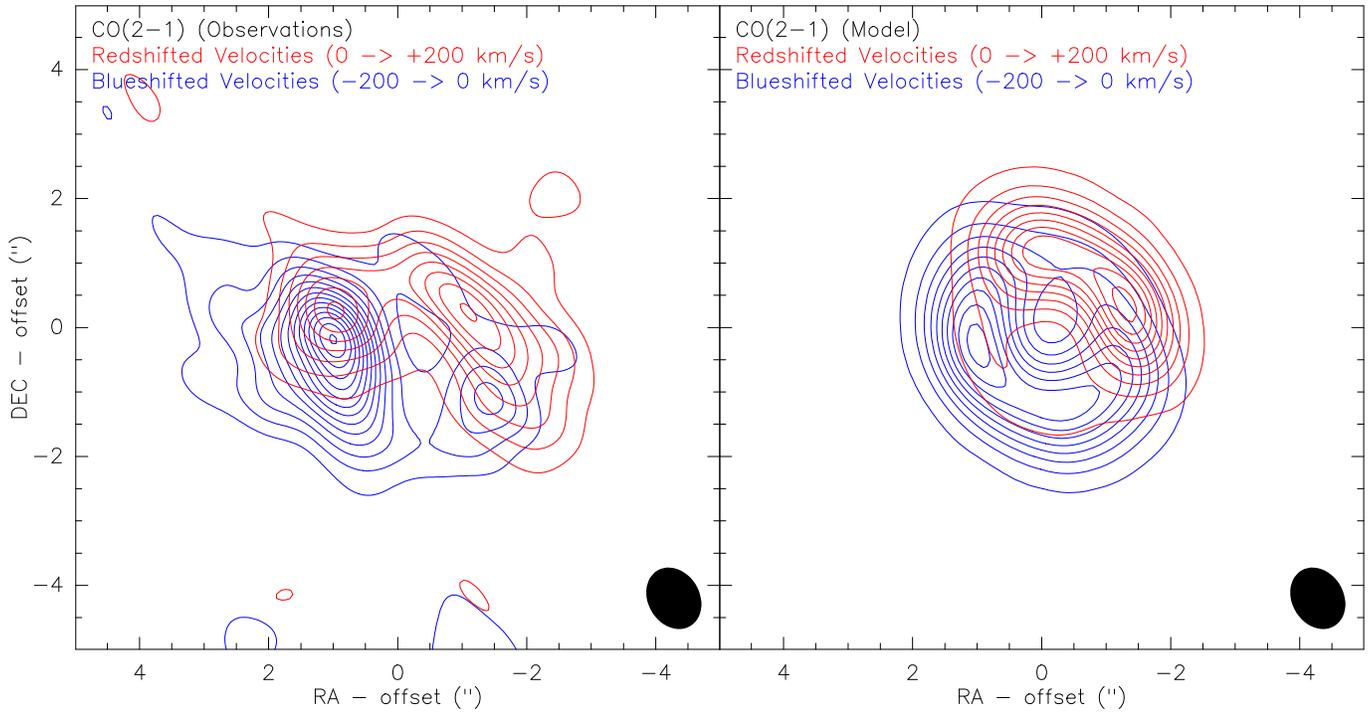}}}
\caption{Blue- and Redshifted emission of the CO model compared to the
$^{12}$CO(2--1) emission.}
\label{fig18.0}
\end{figure*}

\begin{figure*}[!t]
\centering
\rotatebox{-90}{\resizebox{!}{\hsize}{\includegraphics{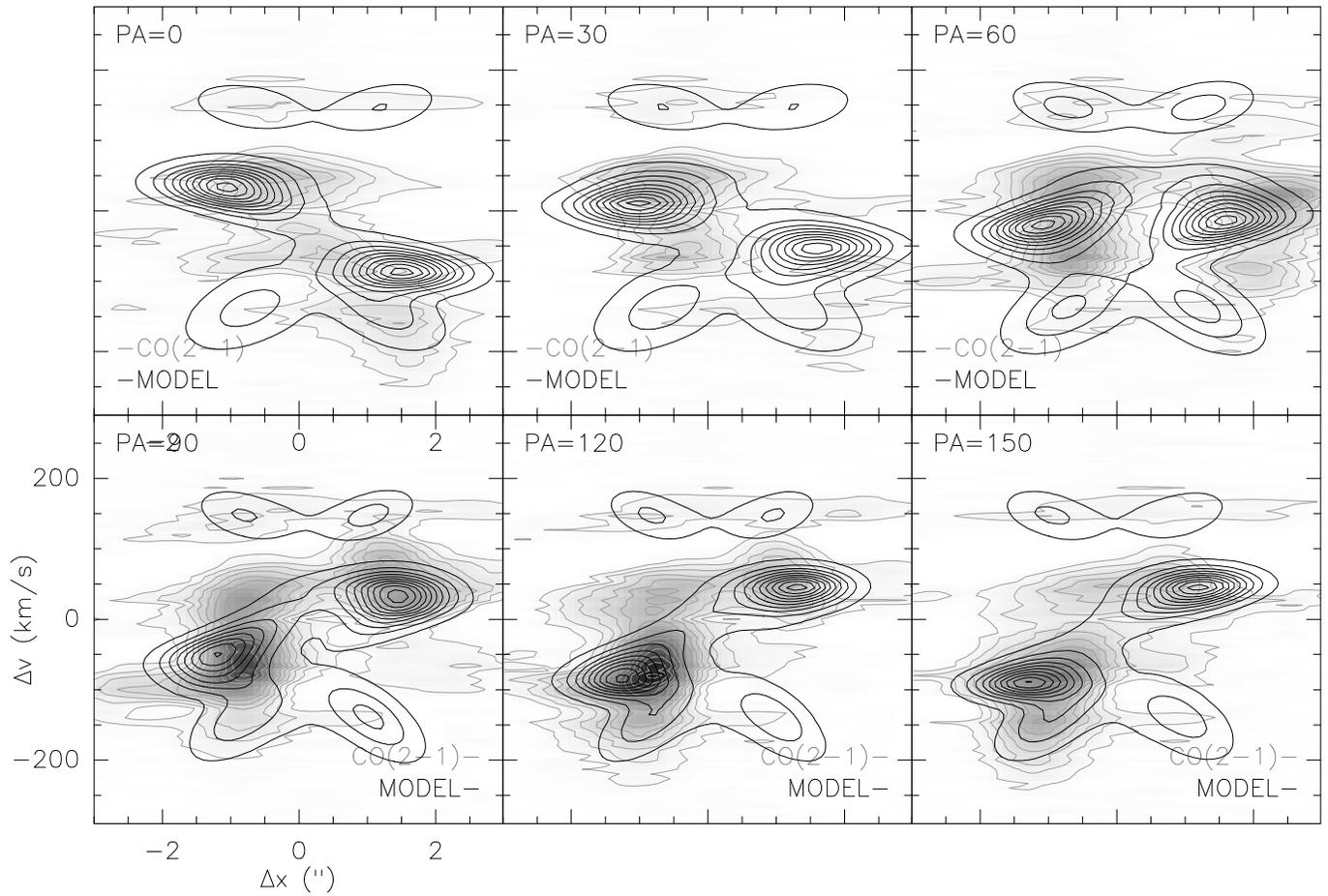}}}
\caption{Position-velocity diagram of the CO model compared to the
$^{12}$CO(2--1) emission.}
\label{fig19.0}
\end{figure*}

Based on this alternative approach, we tried to reproduce the
molecular gas dynamics with a very simplistic model that includes a
dominant (Keplerian) rotating disk plus an outflow of some of the disk
gas. We thereby assume a velocity gradient of $\Delta$v$_0$=200~\kms,
a radius of $\sim$1.2$''$ and an inclination of $\sim$60$^\circ$ for
the rotating disk. We additionally add a slight ellipticity of 5\% and
assymmetry ratio between the eastern and western part of 1:0.7 to the
model. The outflow is approximated by a slightly expanding elliptical
ring. We assume the same ellipticity of 5\% as before and that the
expansion starts at the inner radius of the disk. The expansion rate
(H) is chosen to be 200~\kms\, per 1$''$ or 67~pc equivalently (i.e.,
H=3~km/s/pc); this value is similar to what has been estimated for M51
\citep[2.2~km/s/pc;][]{mats07}. We also introduced a slight asymmetry
ratio of 1:0.9 between the blue- and redshifted emission.  We further
assume that up to $\sim$30\% of the disk gas is expanding. Most of
these values that were chosen for the model parameters were almost
instantly obvious from the observations, especially the velocities and
radii. We hence used them as starting values and scanned then through
a reasonable parameter space for the optimal combination of input
values that matches best the observed maps.  However, given the larger
number of parameters used in this model and hence the many degrees of
freedom, we did not actually conduct a true fit to the data but rather
a ``fit'' by eye. The results of this so found ``best-fit'' model are
displayed and compared to the $^{12}$CO(J=2--1) emission in
Fig.~\ref{fig15.0}-\ref{fig19.0}. Indeed, the major dynamical
characterstics of the molecular gas emission can be reproduced by this
simplistic model supporting the hypothesis of an additional gas
outflow. Also, using the same kinematic parametrisation for the model
as used for the $^{12}$CO emission (see Fig.~\ref{fig10.5}), we find
very similar ratios between the observed and simulated emission. We
have to stress that this suggestion does not exclude a warped CND but
it is not needed for this model.

It is interesting to note as further support for our approach that the
distribution of the HCN(J=3--2) emission matches almost exactly that
of H$_2$ 1--0 S(1) in the CND, even better than the $^{12}$CO(J=2--1)
emission \citep[see Fig.~1 in][]{mueller09}. Similar to the H$_2$ map
that indicates the brightest emission toward the North, also most of
the HCN emission is found toward the northern part of the CND at which
most of the potentially shocked gas would lie. Thus, one would expect
the densest (and hottest) part of the gas in that area (see next
section).

\subsection{Excitation conditions of the gas}

The line ratios derived from the interferometric maps, especially for
the HCN, HCO$^+$ and $^{12}$CO emission (Fig.~\ref{fig11.0} and
Table~\ref{tab6}), are consistent with previous findings from
single-dish observations \citep[e.g.,][]{krips08}. They support a
picture in which the molecular gas in the CND is relatively dense
(n(H$_2$)$\leq$10$^{4.5}$) and warm (T$_{\rm k}$$>$40~K) with
potentially higher than normal (i.e., in galactic giant molecular
clouds) HCN abundances (Z(HCN)=[HCN]/[H$_2$]) of
Z(HCN)=1-50$\times$Z$_{\rm galactic}$(HCN) \citep[Z$_{\rm
galactic}$(HCN)=2$\times$10$^{-8}$; e.g.,][]{irvi87}.  \cite{krips08}
argue that the high HCN-to-$^{12}$CO(J=1--0) and
HCN-to-$^{13}$CO(J=1--0) in NGC~1068 can be explained by either higher
than normal HCN abundances due to an XDR \citep[see
also][]{user04,ster94} and/or higher gas temperatures leading to
hypo-excited CO(J=1--0) emission.  The latter is supported by
decreasing HCN-to-CO line ratios with increasing rotational number J.
However, strong constraints on the kinetic temperatures could not be
set based on the single-dish observations alone. Also, most
single-dish observations are unable to unambigiously distinguish
between the molecular gas emission in the center and that in the
star-forming spiral arms, complicating any interpretation of the
data. Furthermore, most of the interferometric data previously
published mostly focus on $^{12}$CO at moderate angular resolution and
only two of its transitions.

The new interferometric maps, obtained for several transitions and
molecules at sufficiently high angular resolution, overcome some of
the short-comings of previous observations/analyses.

In the following we will discuss results from simulations of the
excitation conditions of the molecular gas carried out with the
radiative transfer code RADEX developed by \cite{vdtak07}.  Please
note that we did not find significant difference when using the LVG
code in MIRIAD or the RADEX code. Given simplified simulations with
RADEX, we decided to use RADEX in this paper as opposed to
\cite{krips08} in which an LVG code was used.

RADEX offers three different possibilities for the escape probability
method: a) a uniform sphere, b) an expanding sphere
(Large-Velocity-Gradient, LVG), and c) a plane parallel slab (shock).
We used all three methods but did not find significant differences for
our data between them.  Thus, in order to keep the interpretation as
simple as possible, we will discuss the results with respect to a
uniform sphere in the following.

We carried out simulations with RADEX for each molecule using the
following grid of parameters (dimension: 51$\times$51$\times$51):

\begin{itemize}

\item Kinetic Temperature:
 T$_{\rm kin}$= 1 - 500~K 
  
\item Gas density:
 n(H$_2$)=10$^3$ - 10$^{7}$~cm$^{-3}$

\item Column density:
 N($^{12}$CO) = 10$^{13}$ - 10$^{21}$~cm$^{-2}$

\end{itemize}

We define the abundance ratio between one molecule (MOL1) and another
(MOL2) as X$_{\rm MOL2}^{\rm MOL1}$$\equiv$Z(MOL1)/Z(MOL2) with
Z(MOL1)$\equiv$[MOL1]/[H$_2$].  For the different molecular abundance
ratios we assume the following ranges:

\begin{itemize}
\item Abundance ratios:\\[0.1cm]
\begin{tabular}{lll}
X$_{\rm ^{13}CO}^{\rm ^{12}CO}$ & $\simeq$ & 6-1000  \\[0.07cm]
X$_{\rm HCO^+}^{\rm HCN}$ & $\simeq$ & 1-500    \\[0.07cm]
\end{tabular}
\end{itemize}

The ranges were chosen such that they span the abundance ratios found
in different (galactic and extragalactic) environments. In the Milky
Way typical values of the $^{12}$C/$^{13}$C abundance ratio are
$\sim$20 which increase to values of 80-100 in the outer parts of the
galaxy \citep[e.g.][]{wil94,wil92}. Nearby starburst galaxies and
ULIRGs show somewhat higher $^{12}$C/$^{13}$C abundance ratios of
$>$30 than found in the Galactic Center \citep[e.g., NGC~253, M82,
IC~342, NGC~4945, NGC~6240 see][]{grev09,henk98,hen94,hena93,henb93}
with Arp~220 being the exception. \cite{grev09} find a very low
abundance ratio of only 8 for this galaxy. Furthermore, some high
redshift galaxies seem to exhibit also rather high $^{12}$C/$^{13}$C
values of $>$30 as determined for the ISM in the graviational lens of
PKS~1830$-$211 \citep{mul06} or the Cloverleaf \citep{henk10}. Values
of X$_{\rm HCO^+}^{\rm HCN}$ show an equally large scatter ranging from
the order of unity in starforming regions in the Milky Way \citep[such
as Orion and SgrB2; see][]{bla87} and nearby starburst galaxies/ULIRGs
\citep[such as M82, NGC~2146, NGC~253, NGC~4945, NGC~6240 and Arp~220;
see][]{nay10,grev09,krips08,wan04} to $\geq$10 in nearby Seyfert
galaxies \citep[e.g.,][]{krips08}.

For the simulations, we will concentrate on the region of the CND that
contains all molecules. This region corresponds to the bridge between
the eastern and western knot, i.e., the northern part of the CND, and
has a size of roughly $\sim$2-3$''$ ($\simeq$120-180~pc). This
corresponds to roughly up to 50\% of the molecular gas in the CND. If
jet interaction indeed plays a role in NGC~1068, this will be the
region most obviously affected.

\subsubsection{$^{12}$CO \& $^{13}$CO emission}
\begin{figure*}[!t]
\centering
\rotatebox{-90}{\resizebox{!}{\hsize}{\includegraphics{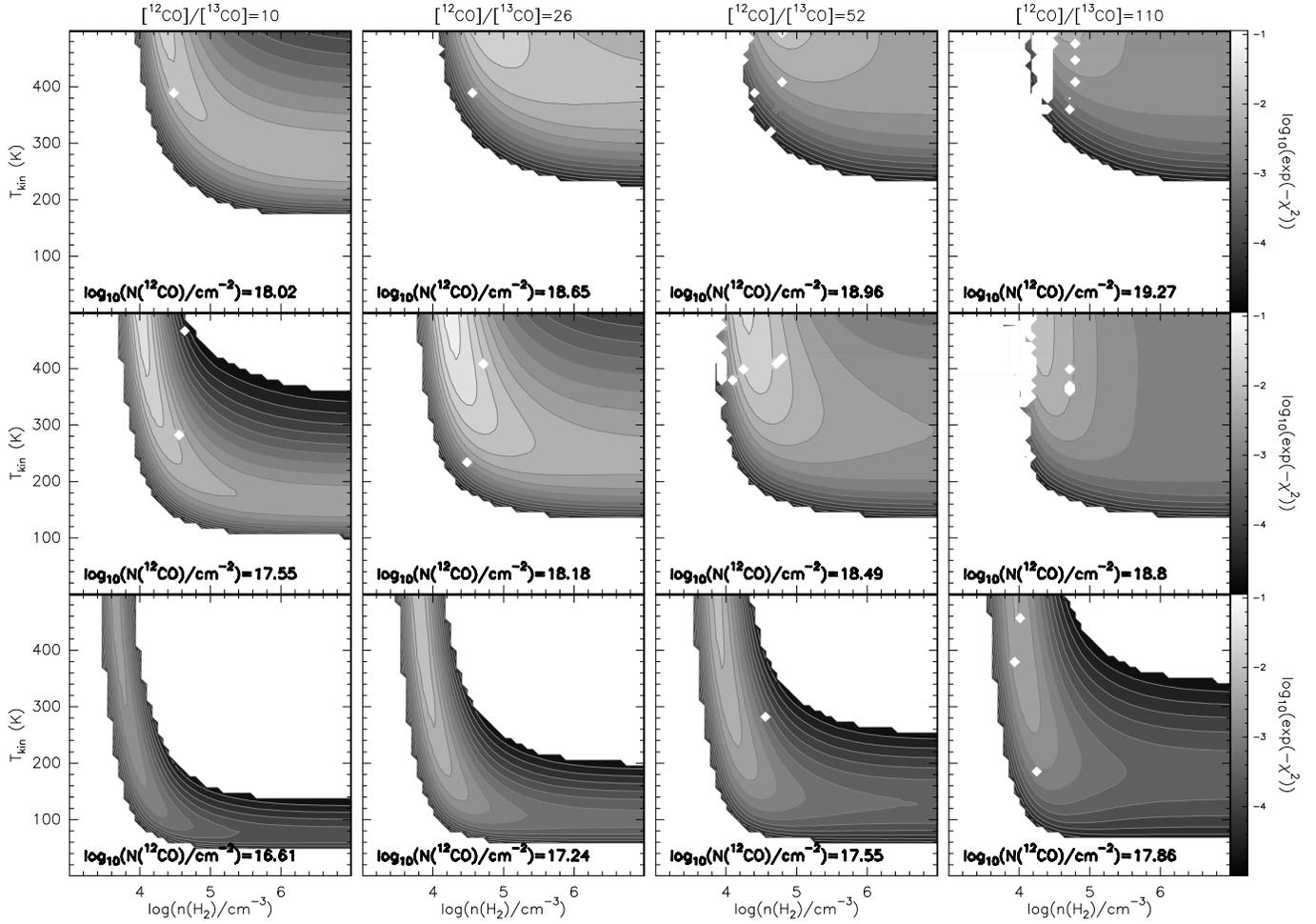}}}
\caption{$\chi^2$-fit results obtained from the RADEX simulations of
 the excitation conditions of the molecular gas. Shown are four
 different [$^{12}$CO]/[$^{13}$CO] abundance ratios (=10,26,52,110)
 for three different $^{12}$CO column densities respectively. The
 middle panel shows the best $\chi^2$-fit found for each abundance
 ratio.}
\label{fig20.0}
\end{figure*}

We conducted a $\chi^2$-test on the RADEX grid by using the observed
line ratios for the $^{12}$CO and $^{13}$CO line
emission. Fig.~\ref{fig20.0} shows the parameters for the best
$\chi^2$-test for four exemplary abundance ratios (10,26,52,110) from
the aforementioned range; abundance ratios at the lower and higher end
of the range show somewhat higher $\chi^2$ values.

The middle panel shows the lowest $\chi^2$ found in each range of
column densities and abundance ratios. The lower and upper panels show
the respective lower and upper limits for the column density for which
still a reasonably low $\chi^2$ was found to indicate the spread in
column densities.

The availability of the three lowest transitions for $^{12}$CO and
$^{13}$CO allows us to set tight constraints on T$_{\rm kin}$,
n(H$_2$) and N($^{12}$CO).  The observed CO line ratios most
impressively restrict the kinetic temperatures to values well above
200~K. This strengthens previous indications of warm/hot ( T$_{\rm
kin}$$>$50~K) molecular gas in the CND of NGC~1068
\citep{kame11,krips08,mats98,ster94} but being much higher than the
value found by \cite{tac94}. However, \cite{tac94} base their
simulations on single-dish data which cannot distinguish well between
emission from the starforming ring/spiral arms and the CND. Especially
the $^{13}$CO(J=1--0) emission might be overestimated for the CND
which clearly is hardly detected in the interferometric map despite
the high sensitivity of the observations.  The lowest $\chi^2$ is
actually found for kinetic temperatures around 450~K, which seems to
be fairly high. Although we used a one-gas component model due to the
lack of sufficient observational constraints (i.e., higher-J CO
transitions with J$_{\rm upper}$$\geq$4), we do not expect all of the
gas to be at these high kinetic temperatures, but rather a fraction of
the gas in the northern part of the bridge. It seems likely that by
using a two-temperature model for the molecular gas, we may find
slightly lower maximum temperatures. However, this necessitates either
the inclusion of CO data at higher-J transitions or using so called
molecular thermometers \citep[such as H$_2$CO or NH$_3$;
see][]{ao11}. High kinetic temperatures would be expected in several
scenarios, among them shocks as well as heating through X-ray
radiation from the AGN \citep[e.g.,][]{meij07,meij05}. Nevertheless, a
significant fraction ($>>$10-20\%) of the warm molecular gas in this
bridge regions seems to exhibit high kinetic temperatures which is a
much larger fraction than so far found (for the overall molecular gas
reservoir) in starburst and other Seyfert galaxies \cite[$\leq$30\%,
see for instance][]{rous07,dale05,rigo02}.

The density is also well constrained and is found to be in the range
of 10$^{3.5}$-10$^{4.5}$~cm$^{-3}$, consistent with previous findings
\citep{krips08,mats98}. Considering the range of assumed abundance
ratios, the column density of $^{12}$CO approximately spans a range
between $\sim$10$^{17.0}$~cm$^{-2}$ and 10$^{19.0}$~cm$^{-2}$. Even
though CO might not be the best tracer of the $^{12}$C/$^{13}$C
isotopic ratio \citep[e.g.,][]{mar10} we assume it to be a first
approximation to this ratio.  The carbon ratio of 26 found for the
absolute lowest $\chi^2$ is similar to the value of 20 measured in the
Galactic Center region \citep[e.g.,][]{wil94} and lower than that
derived towards nearby starbursts
\citep[e.g.,][]{hena93,henb93,hen94}.  Such $^{13}$C enrichment would
point towards a highly nuclear processing of the ISM in the central
region of NGC~1068.

\subsubsection{HCN \& HCO$^+$ emission}

\begin{figure*}[!t]
\centering
\rotatebox{-90}{\resizebox{!}{\hsize}{\includegraphics{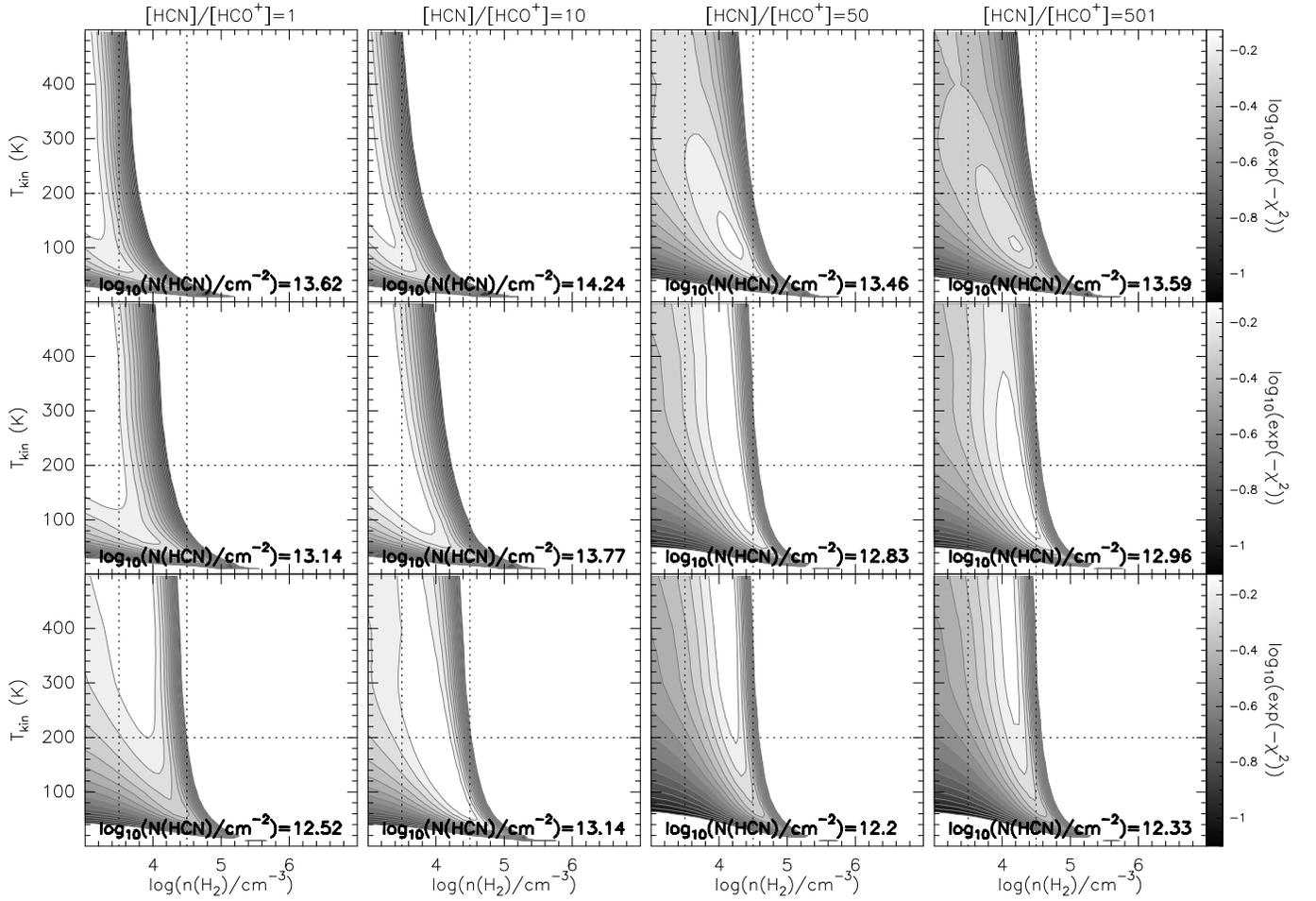}}}
\caption{$\chi^2$-fit results obtained from the RADEX simulations of
 the excitation conditions of the molecular gas. Shown are four
 different [HCN]/[HCO$^+$] abundance ratios around the standard
 galactic value of [HCN]/[HCO$^+$]$\simeq$10 for three different HCN
 column densities respectively. The middle panel shows the best
 $\chi^2$-fit found for each abundance ratio.}
\label{fig21.0}
\end{figure*}

In \cite{krips08} we carried out an LVG analysis for NGC~1068 based on
the HCN and HCO$^+$ single-dish lines ratios with kinetic temperatures
not exceeding 200~K. As our new simulations indicate kinetic
temperatures lying significantly above 200~K, we repeated the
simulations with RADEX allowing for a larger range in kinetic
temperatures.  The line fluxes for the HCN and HCO$^+$(J=1--0)
emission are taken from PdBI observations at $\sim$1$''$ angular
resolution \citep[][]{garc08} which will be analysed in more detail in
a later paper by Usero et al.\ (in prep.). The results of the RADEX
simulations for HCN and HCO$^+$ are shown in Fig.~\ref{fig21.0}.

Considering the restrictions for the kinetic temperatures ($>$200~K),
we find solutions (i.e., with low $\chi^2$) with RADEX for which the
HCN-to-HCO$^+$ abundance ratio lies between X$^{\rm HCN}_{\rm
HCO^+}$$\simeq$10-500 which is higher than that found in
starforming/starbursting regions (X$^{\rm HCN,g}_{\rm
HCO^+}$$\simeq$1). However, good solutions are also found for lower
kinetic temperatures. This is due to the fact that HCN and HCO$^+$ are
not as sensitive to changes in the kinetic temperatures as $^{12}$CO
and $^{13}$CO. They are better indicators of changes in the volume
density, as can be nicely seen in Fig.~\ref{fig21.0}; the volume
density is restricted by the $\chi^2$-test to a very small area and
independently yields values of
n(H$_2$)$\simeq$10$^{3.5}$-10$^{4.5}$~cm$^{-3}$ similar to the
$^{12}$CO and $^{13}$CO results.

The simulations\footnote{Please note that we consider the two highest
contours in Fig.~\ref{fig21.0} as being acceptable solutions} indicate
column densities for HCN in the range of
N(HCN)$\simeq$10$^{12.0}$-10$^{13.5}$~cm$^{-2}$ which is smaller (a
factor of $\sim$10) than what was found by \cite{krips08} with the LVG
code. However, results are quite similar to \cite{krips08} when
assuming similar kinetic temperatures.

Comparing the HCN column densities to those of $^{12}$CO
(N($^{12}CO)$$\simeq$10$^{17.0}$-10$^{19.0}$~cm$^{-2}$), we obtain
abundance ratios between HCN and $^{12}$CO of X$^{\rm ^{12}CO}_{\rm
  HCN}$$\gtrsim$10$^{3.5}$ which seems to be still compatible with a
slightly increased abundance of HCN. Comparing the column densities of
HCO$^+$ and $^{12}$CO, we find a somewhat decreased HCO$^+$ abundance
(by a factor of at least 10 lower than found in galactic starforming
regions).  This agrees well with previous results
\citep[e.g.,][]{garc10,krips08,user04} suggesting an increased
formation (and hence increased abundance) of HCN due to an XDR in the
center of NGC~1068.

\section{Summary \& Conclusions}
\label{sum}

The SMA and PdBI observations of the (sub-)mm emission in NGC~1068
presented in this paper show a complex distribution, kinematics and
excitation conditions of the molecular gas. The (sub)mm continuum and
molecular line emission is interpreted as follows:

\begin{itemize}
\item[1.)] The cm/mm-continuum emission seems to be best reproduced by
           electron-scattered synchrotron emission. Thermal free-free
           emission as proposed by \cite{hoen08} overpredicts the high
           angular resolution 1~mm continuum emission.

\item[2.)] The molecular gas is found to display a very complex
           kinematic behaviour in the $^{12}$CO, HCN and HCO$^+$ lines
           which is not reproducable by a tilted-ring model
           approximating a warped disk with circular motions.
           Instead, a dominant rotating disk plus a radial outflow of
           some of the gas in the CND is proposed as an alternative
           explanation to account for the non-circular motions.
  
\item[3.)] The different line ratios from the $^{12}$CO, $^{13}$CO,
           HCN, and HCO$^+$ emission seems to be consistent with
           moderately dense and warm gas, both being further support
           for a gas scenario in which heated and compressed by a
           shock (at least in the North/North-Eastern part of the
           ring). The highest line ratios are found close to the AGN
           and/or jet-CND 'contact'-point. In this picture, the
           increased kinetic temperatures seem to be one of the
           culprits for the unusually high HCN-to-CO(J=1--0) line
           ratios due to a hypo-excitation of the CO(J=1--0) line
           emission.

\item[5.)] Consistent with previous papers, we find further
           indications of an increased HCN abundance in NGC~1068 (by a
           factor of $\sim$4-10), and a decreased HCO$^+$ abundance
           (by a factor of $\sim$5-10), explaining the high
           HCN-to-HCO$^+$ abundance ratio in the CND of this source.

\end{itemize}

\acknowledgments 

The Submillimeter Array is a joint project between the Smithsonian
Astrophysical Observatory and the Academia Sinica Institute of
Astronomy and Astrophysics and is funded by the Smithsonian
Institution and the Academia Sinica. We thank the anonymous referee
for a very careful and constructive report. We also would like to
thank Gaelle Dumas for very useful discussion on and help with the
kinematic analysis of the $^{12}$CO data. I.S. acknowledges grant
LC06014 of Ministry of Education of the Czech Republic.  ABP thanks
the National Radio Astronomy Observatory (NRAO) for support on this
project.  NRAO is a facility of the National Science Foundation
operated under cooperative agreement by Associated Universities, Inc.
SM is supported by the National Science Council (NSC) of Taiwan, NSC
97-2112-M-001-021-MY3.

\begin{deluxetable}{ccc}
\tablecolumns{3}
\tabletypesize{\small} 
\tablewidth{\hsize} 
\tablecaption{Properties of NGC~1068 \label{tab1}}
\tablehead{Characteristic & NGC~1068 & Reference}
\startdata 
Hubble Class\dotfill & (R)SA(rs)b & NED$^a$\\
AGN type\dotfill            & Seyfert~2 & \cite{khac74} \\
Dynamical Center\dotfill     & $\alpha_{\rm J2000}$=02h42m40.70s 
 & \multirow{2}{*}{\cite{gall04}} \\
                     & $\delta_{\rm J2000}$=$-$00$^\circ$00$'$$47\farcs9$ &  \\
Redshift\dotfill              & 0.00379$\pm$0.00001   & \cite{huch99} \\
Systemic Velocity\dotfill     & (1137$\pm$3)\kms & \cite{huch99} \\
Luminosity Distanc\dotfill e  & 12.6~Mpc$^b$ & NED$^a$ \\
Scale\dotfill                 & 61~pc/arcsec & NED$^{a,b}$ \\
Inclination Angle\dotfill     & 40$^\circ$   & \cite{blan97} \\ 
Position Angle\dotfill        & 278$^\circ$  & \cite{blan97}  \\
\enddata 
\tablenotetext{a}{NED: NASA/IPAC Extragalactic Database}
\tablenotetext{b}{Used cosmology: Hubble constant H$_0$ =
  73~\kms/Mpc, Omega(matter) = 0.27, Omega(vacuum) = 0.73}
\end{deluxetable}

\begin{deluxetable}{lccccccccc}
\tablecolumns{10}
\tabletypesize{\small} 
\tablewidth{\hsize} 
\tablecaption{Chronological Summary of observations carried out for NGC~1068\label{tab2}}
\tablehead{
\colhead{Molecular} & \colhead{Telescope} & \colhead{Observing} & \colhead{Config.$^a$} & \colhead{Frequency} & \colhead{Band} 
& \colhead{Zenith}     & \colhead{T$_{\rm sys}$} & \colhead{RMS} & \colhead{Synthesized Beam} \\
\colhead{Line}      & \colhead{}          & \colhead{Dates}     & \colhead{}            & \colhead{at rest}   & \colhead{}     
& \colhead{Opacity}    &  \colhead{}             & \colhead{Noise$^b$}  & \colhead{major$\times$minor,P.A.$^c$}\\
\colhead{}          & \colhead{}          & \colhead{(YYYY-MM)} &  \colhead{}           & \colhead{(GHz)}     &  \colhead{}    
& \colhead{at 225~GHz} & \colhead{(K)}           & \colhead{(mJy)} & \colhead{($''$$\times$$''$,$^\circ$)} }
\startdata 
$^{13}$CO(J=2--1) & SMA   & 2008-01,2008-02 & EX  & 220.399 & LSB & 0.10-0.20 & 100-150 & 12     & 1.0$\times$0.8,30$^d$ \\
\, \& C$^{18}$O(J=2--1) & &                 &     & 219.560 & LSB &           &         &        &                       \\			      	 				 			 
HCO$^+$(J=4--3)   & SMA   & 2007-08,2007-09 & EX  & 356.734 & USB & 0.06-0.13 & 200-400 & 32$^e$ & 1.0$\times$0.8,30$^f$ \\
$^{12}$CO(J=3--2) & SMA   & 2007-09         & EX  & 345.796 & LSB & 0.06-0.13 & 200-400 & 51     & 1.0$\times$0.8,30$^f$ \\
HCO$^+$(J=3--2)   & SMA   & 2006-11         & EX  & 267.558 & LSB & 0.06-0.15 & 100-150 & 33     & 1.0$\times$0.8,30$^f$ \\
HCN(J=3--2)       & SMA   & 2006-01         & EX  & 265.886 & LSB & 0.05-0.10 & 100-150 & 24$^e$ & 1.0$\times$0.8,30$^{e,f}$ \\
                  & SMA   & 2006-10,2006-11 & VEX &         & LSB & 0.10-0.22 & 100-200 &        & 0.53$\times$0.46,30$^{d,e}$ \\              
$^{13}$CO(J=3--2) & SMA   & 2005-10         & C   & 356.734 & USB & 0.05-0.06 & 200-300 & 39     & 2.4$\times$2.1,28$^d$ \\ 
$^{13}$CO(J=1--0) & PdBI  & 2003-02         & A   & 110.201 & SSB & n.a.      & 150-300 & 1.7    & 2.5$\times$1.9,28$^f$ \\
$^{12}$CO(J=2--1) & PdBI  & 2003-02         & A   & 230.538 & SSB & n.a.      & 200-600 & 4      & 1.0$\times$0.8,30$^e$ \\[-0.15cm]
\enddata 
\tablenotetext{a}{SMA configurations: C=compact (baselines up to
70~m), EX=extended (baselines up to 220~m), VEX=very extended
(baselines up to 500~m); PdBI configurations: A=most extended
(baselines up to 500m)}
\tablenotetext{b}{in 17~\kms\ wide channels}
\tablenotetext{c}{P.A.\ is measured from North to East}
\tablenotetext{d}{using uniform weighting}
\tablenotetext{e}{combined for all tracks}
\tablenotetext{f}{using natural weighting}

\end{deluxetable}

\begin{deluxetable}{ccccccc}
\tabletypesize{\small} 
\tablecaption{Continuum Parameters for NGC~1068.}
\tablewidth{\hsize} 
\tablehead{
$\lambda$ & Synth.\ Beam & $\Delta\alpha$$^a$ & $\Delta\delta$$^a$ & Peak & Flux        
 & Deconv.\ Size\\
  & major$\times$minor,P.A. &  &  & Flux & Density$^b$ & major$\times$minor,P.A. \\
  & ($''$$\times$$''$,$^\circ$)  & ($''$) & ($''$) & (mJy/beam) & (mJy) 
&  ($''$$\times$$''$,$^\circ$) }
\startdata 
1.4~mm   & 1.0$\times$0.8,30   & $+$0.13$\pm$0.02 & $+$0.25$\pm$0.02 & 19$\pm$2 & 28$\pm$3
    & (0.9$\pm$0.1)$\times$(0.7$\pm$0.1),(50$\pm$20) \\
1.3~mm   & 1.0$\times$0.8,30   & $+$0.18$\pm$0.02 & $+$0.23$\pm$0.02 & 15$\pm$1 & 22$\pm$2 
    & (0.6$\pm$0.1)$\times$(0.5$\pm$0.1),(40$\pm$20) \\
1.0~mm (NA)$^c$  & 1.0$\times$0.8,30   & $+$0.17$\pm$0.05 & $+$0.15$\pm$0.10 & 19$\pm$2 & 24$\pm$3 
    & (0.8$\pm$0.1)$\times$(0.4$\pm$0.1),(20$\pm$10) \\
1.0~mm (UN)$^d$  & 0.5$\times$0.4,30   & $+$0.13$\pm$0.03 & $+$0.07$\pm$0.03 & 12$\pm$2 & 13$\pm$2 
    & (0.3$\pm$0.1)$^e$ \\
810~$\mu$m  & 2.1$\times$2.0,80 & $-$0.04$\pm$0.2 & $-$0.07$\pm$0.2 & 30$\pm$4 & 41$\pm$11
    & (1.1$\pm$0.1)$\times$(0.8$\pm$0.2),(90$\pm$20)\\
850~$\mu$m  (NA) & 1.0$\times$0.8,30 & $+$0.33$\pm$0.07 & $+$0.05$\pm$0.06 & 24$\pm$2 & 50$\pm$7
    & (1.1$\pm$0.1)$\times$(0.8$\pm$0.2),(90$\pm$20)\\
850~$\mu$m  (UN) & 0.6$\times$0.5,30 & $+$0.33$\pm$0.09 & $-$0.03$\pm$0.07 & 16$\pm$3 & 30$\pm$5
    & (1.1$\pm$0.1)$\times$(0.8$\pm$0.2),(90$\pm$20)\\[-0.15cm]
\enddata 

\tablenotetext{$a$}{The offsets are with respect to $\alpha_{\rm
  J2000}=$02h42m40.70s and $\delta_{\rm
  J2000}=$-00$^\circ$00$'$$47\farcs9$ which is almost identical to the
  radio position of the AGN in NGC~1068 of $\alpha_{\rm
  J2000}=$02h42m40.709s and $\delta_{\rm
  J2000}=$-00$^\circ$00$'$$47\farcs95$
  \citep[e.g.][]{gall04,krip06}. Positional errors are of pure
  statistical nature and were derived from the Gaussian fit to the
  data. They do not include absolute positional uncertainties from the
  calibration, which are estimated to be $\sim$$0\farcs1$. }

\tablenotetext{$b$}{Flux errors are purely statistical and do not
  account for uncertainties of the flux calibration. The latter are
  estimated to be of the order 10-20\% (see text).}

\tablenotetext{$c$}{Averaged continuum emission derived from the HCN(J=3--2)
  (vex+ext) and HCO$^+$(J=3--2) observations (ext). Data were mapped using
  natural weighting (NA).}

\tablenotetext{$d$}{Averaged continuum emission derived from the
  HCN(J=3--2) observations alone (vex+ext) using uniform weighting
  (UN).}

\tablenotetext{$e$}{Here, only a circular Gaussian fit has been
carried out, while for the rest an elliptical Gaussian has been fitted
to the data (see text).}

\label{tab3}
\end{deluxetable}

\begin{deluxetable}{ccccccc}
\tabletypesize{\small} 
\tablecaption{Individual components of the molecular line emission in NGC~1068.}
\tablewidth{\hsize} 
\tablehead{
Molecular & Component & $\Delta\alpha$$^{a}$ & $\Delta\delta$$^{a}$ & Vel.\ Integrated
 & Spatially Integrated \\
Line      &           &                &                & Peak Intensity$^{a}$ & Intensity$^{a}$   \\
          &           & ($''$)         & ($''$)         & (Jy~beam$^{-1}$\kms) & (Jy~\kms)   }
\startdata 

HCN(J=3--2)       & E-knot & $+$1.0$\pm$0.1 & $+$0.1$\pm$0.1 &  51$\pm$8   &  110$\pm$20  \\
                  & W-knot & $-$0.8$\pm$0.2 & $+$0.2$\pm$0.2 &  32$\pm$4   &   70$\pm$10  \\
HCO$^+$(J=3--2)   & E-knot & $+$0.9$\pm$0.1 & $+$0.0$\pm$0.1 &  28$\pm$5   &   52$\pm$5   \\
                  & W-knot & $-$0.8$\pm$0.2 & $+$0.5$\pm$0.2 &  14$\pm$2   &   40$\pm$5   \\
HCO$^+$(J=4--3)   & E-knot & $+$0.9$\pm$0.1 & $+$0.1$\pm$0.1 &  27$\pm$6   &   98$\pm$20  \\
$^{12}$CO(J=3--2) & E-knot & $+$1.1$\pm$0.1 & $-$0.1$\pm$0.1 & 470$\pm$20  & 1330$\pm$200 \\
                  & W-knot & $-$1.2$\pm$0.1 & $-$0.1$\pm$0.1 & 270$\pm$10  &  720$\pm$100 \\
$^{13}$CO(J=3--2) & E-knot & $+$1.1$\pm$0.1 & $+$0.3$\pm$0.1 &  24$\pm$3   &   40$\pm$10  \\
$^{12}$CO(J=2--1) & E-knot & $+$1.0$\pm$0.1 & $-$0.2$\pm$0.1 &  70$\pm$10  &  290$\pm$20  \\
                  & W-knot & $-$1.3$\pm$0.1 & $-$0.3$\pm$0.1 &  30$\pm$5   &  180$\pm$20  \\
$^{13}$CO(J=2--1) & E-knot & $+$0.6$\pm$0.2 & $+$0.4$\pm$0.2 & 10$\pm$2    &   18$\pm$3   \\ 
                  & W-knot & $-$1.2$\pm$0.3 & $+$0.1$\pm$0.1 & 8$\pm$2     &   16$\pm$3   \\
C$^{18}$O(J=2--1) & E-knot & $+$0.3$\pm$0.3 & $+$0.5$\pm$0.3 & 7$\pm$1     &   15$\pm$3   \\ 
$^{13}$CO(J=1--0) & E-knot & $+$1.0$\pm$0.4 & $-$0.4$\pm$0.4 & 0.4$\pm$0.2 &  0.5$\pm$0.2 \\ 
                  & W-knot & $-$1.3$\pm$0.4 & $-$0.8$\pm$0.5 & 0.4$\pm$0.2 &  0.5$\pm$0.2 \\
$^{12}$CO(J=1--0) & E-knot & $+$0.5$\pm$0.1 & $+$0.1$\pm$0.1 & 40$\pm$1    &   90$\pm$2   \\ 
                  & W-knot & $-$1.1$\pm$0.1 & $-$0.5$\pm$0.1 & 19$\pm$1    &   40$\pm$2   \\[-0.15cm]
\enddata 

\tablenotetext{$a$}{The parameters were determined by fitting a one-
  or two-component elliptical Gaussian profile to the $uv$-data of
  each line. Errors include the statistical uncertainties from the
  Gaussian fit and those from the calibration ($\sim$10-20\%). Offsets
  are with respect to the center position specified in
  Table~\ref{tab3}.}

\label{tab4}
\end{deluxetable}

\begin{deluxetable}{ccccccc}
\tabletypesize{\small} 
\tablecaption{Molecular line parameters derived from the different line spectra
  of NGC~1068.}
\tablewidth{\hsize} 
\tablehead{
Molecular & Velocity   &  Line     & Line      & Vel.\ Integrated    & Vel.\ Integrated        & SD Beam\\
Line      & Offset$^{a,b,c}$ & Flux$^{a}$ & Width$^{a,b,d}$ & Intensity$^{a,b}$ & SD Intensity$^e$&($``$)\\
          & (\kms)     & (Jy)    & (\kms)    & (Jy~\kms) & (Jy~\kms)
  }
\startdata 
HCN(J=3--2)       & $-$30$\pm$10 & 0.63$\pm$0.08   & 200$\pm$30 &  150$\pm$10  &  190$\pm$10  &  9.5$''$ \\
HCO$^+$(J=3--2)   & $-$40$\pm$10 & 0.23$\pm$0.03   & 190$\pm$50 &   50$\pm$6   &   80$\pm$8   &  9.5$''$ \\
HCO$^+$(J=4--3)   & $-$40$\pm$10 & 0.24$\pm$0.03   & 240$\pm$40 &   60$\pm$6   &   70$\pm$10  &   14$''$ \\
$^{12}$CO(J=3--2) & $-$30$\pm$3  & 12.3$\pm$0.40   & 170$\pm$10 & 2130$\pm$5   & 2600$\pm$300 &   14$''$ \\
$^{13}$CO(J=3--2) & $-$40$\pm$10 & 0.43$\pm$0.03   & 230$\pm$30 &  100$\pm$10  &  170$\pm$20  &   14$''$ \\
$^{12}$CO(J=2--1) & $-$30$\pm$6  & 2.20$\pm$0.10   & 230$\pm$20 &  529$\pm$2   &  950$\pm$6   &   12$''$ \\
$^{13}$CO(J=2--1) & $-$10$\pm$10 & 0.50$\pm$0.04   &  60$\pm$10 &   30$\pm$1   &   55$\pm$7   &   12$''$ \\
C$^{18}$O(J=2--1) &  $+$3$\pm$10 & 0.10$\pm$0.02   &  50$\pm$10 &  5.1$\pm$0.3 &   \nodata    &  \nodata \\
$^{12}$CO(J=1--0) & $+$3$\pm$2   & 0.50$\pm$0.02   & 240$\pm$10 &  120$\pm$4   &  650$\pm$80  &   21$''$ \\
$^{13}$CO(J=1--0) & $+$8$\pm$10  & 0.009$\pm$0.002 & 140$\pm$30 &  1.3$\pm$0.3 &  56$\pm$7    &   21$''$ \\[-0.15cm]
\enddata 

\tablenotetext{$a$}{The line parameters have been determined by fitting a
  single Gaussian line to the (spatially integrated) spectrum for each
  molecule. The line emission has been thereby integrated over the
  central 4$''$ in NGC~1068.}

\tablenotetext{$b$}{statistical error from the Gaussian fit only.}

\tablenotetext{$c$}{with respect to v$_{\rm
    LSR}$=1137~km/s.}

\tablenotetext{$d$}{Full Width at Half Maximum (FWHM)}

\tablenotetext{$e$}{single dish (SD) integrated intensities as
  measured with the IRAM 30m and the JCMT telescope in the central
  10-30$''$ of NGC~1068 \citep[taken
  from:][]{isra09,perez09,krips08}. The values were converted to
  Jansky scale using S[Jy]/T$_{\rm mb}$[K]=4.71 (30m) and
  S[Jy]/T$_{\rm mb}$[K]=15.6 (JCMT).}

\label{tab5}
\end{deluxetable}

\clearpage

\begin{landscape}
\begin{deluxetable}{rccccccccccccc}
\tabletypesize{\scriptsize} 
\tablecaption{Molecular line ratios for NGC~1068$^a$.  (X[K]/Y[K])}
\tablewidth{0pt} 
\tablehead{ 
 Y  & \multicolumn{3}{c}{$^{12}$CO} & \multicolumn{3}{c}{$^{13}$CO} 
  & \multicolumn{1}{c}{C$^{18}$O} & \multicolumn{3}{c}{HCN} & \multicolumn{2}{c}{HCO$^+$}  \\
  \multicolumn{1}{l}{X}  & \multicolumn{1}{c}{J=1--0} & \multicolumn{1}{c}{J=2--1} & \multicolumn{1}{c}{J=3--2} 
  & \multicolumn{1}{c}{J=1--0} & \multicolumn{1}{c}{J=2--1} & \multicolumn{1}{c}{J=3--2} 
  & \multicolumn{1}{c}{J=2--1} & \multicolumn{1}{c}{J=1--0} & \multicolumn{1}{c}{J=2--1} & \multicolumn{1}{c}{J=3--2} 
  & \multicolumn{1}{c}{J=1--0} & \multicolumn{1}{c}{J=3--2} & \multicolumn{1}{c}{J=4--3}  }
\startdata 
\multicolumn{14}{c}{E-knot}\\
$^{12}$CO J=1--0 & \nodata         &  0.3$\pm$0.2    & 0.2$\pm$0.1     &  20$\pm$10  & \nodata     &  \nodata       & \nodata     & \nodata & \nodata  &  \nodata       & \nodata &  \nodata      & \nodata     \\
          J=2--1 & 2.9$\pm$0.3     &  \nodata        & 0.3$\pm$0.2     &  \nodata    & 5.0$\pm$3.0 &  \nodata       & 17$\pm$4    & \nodata & \nodata  &  4.0$\pm$2.0   & \nodata &  7.0$\pm$3.0  & \nodata     \\
          J=3--2 & 6.0$\pm$3.0     &  4.0$\pm$2.0    & \nodata         &  \nodata    & \nodata     &  25$\pm$6      & \nodata     & \nodata & \nodata  &  5.0$\pm$3.0   & \nodata &  8.0$\pm$3.0  & \nodata     \\
$^{13}$CO J=1--0 & 0.05$\pm$0.03   &  \nodata        & \nodata         &  \nodata    & 0.2$\pm$0.1 &  0.08$\pm$0.04 & \nodata     & \nodata & \nodata  &  \nodata       & \nodata &  \nodata      & \nodata     \\
          J=2--1 & \nodata         &  0.2$\pm$0.1    & \nodata         & 5.0$\pm$3.0 & \nodata     &  0.7$\pm$0.3   & 1.4$\pm$0.3 & \nodata & \nodata  &  0.3$\pm$0.2   & \nodata &  0.5$\pm$0.2  & \nodata     \\
          J=3--2 & \nodata         &  \nodata        & 0.04$\pm$0.01   &  13$\pm$4   & 1.4$\pm$0.4 &  \nodata       & \nodata     & \nodata & \nodata  &  0.2$\pm$0.1   & \nodata &  0.5$\pm$0.3  & \nodata     \\
C$^{18}$O J=2--1 & \nodata         &  0.06$\pm$0.03  & \nodata         &  \nodata    & 0.8$\pm$0.3 &  \nodata       & \nodata     & \nodata & \nodata  &  \nodata       & \nodata &  \nodata      & \nodata     \\
HCN J=1--0       & \nodata         &  \nodata        & \nodata         &  \nodata    & \nodata     &  \nodata       & \nodata     & \nodata & \nodata  &  \nodata       & \nodata &  \nodata      & \nodata     \\
    J=3--2       & \nodata         &  0.3$\pm$0.1    & 0.14$\pm$0.03   &  \nodata    & 4$\pm$1     &  4.0$\pm$1.0   & \nodata     & \nodata & \nodata  &  \nodata       & \nodata &  \nodata      & \nodata     \\
HCO$^+$ J=1--0   & \nodata         &  \nodata        & \nodata         &  \nodata    & \nodata     &  \nodata       & \nodata     & \nodata & \nodata  &  \nodata       & \nodata &  \nodata      & \nodata     \\
        J=3--2   & \nodata         &  0.13$\pm$0.02  & 0.07$\pm$0.01   &  \nodata    & 2.0$\pm$0.4 &  2.0$\pm$0.5   & \nodata     & \nodata & \nodata  &  0.6$\pm$0.3   & \nodata &  \nodata      & 1.8$\pm$0.9 \\
        J=4--3   & \nodata         &  \nodata        & \nodata         &  \nodata    & \nodata     &  \nodata       & \nodata     & \nodata & \nodata  &  \nodata       & \nodata &  0.6$\pm$0.3  & \nodata     \\
\hline
\\[-0.15cm]

\multicolumn{14}{c}{W-knot}\\
$^{12}$CO J=1--0 & \nodata         & 0.8$\pm$0.7     & 0.3$\pm$0.2     & 50$\pm$30   & \nodata       & \nodata      & \nodata     & \nodata & \nodata  & \nodata        & \nodata & \nodata     & \nodata     \\
          J=2--1 & 1.5$\pm$0.5     & \nodata         & 0.3$\pm$0.2     & \nodata     & 10$\pm$2      & \nodata      & \nodata     & \nodata & \nodata  & 3.4$\pm$0.6    & \nodata & 6.0$\pm$1.0 & \nodata     \\
          J=3--2 & 4.0$\pm$2.0     & 3.0$\pm$2.0     & \nodata         & \nodata     & \nodata       & 25$\pm$6     & \nodata     & \nodata & \nodata  & 6.0$\pm$1.0    & \nodata & 11$\pm$2    & \nodata     \\
$^{13}$CO J=1--0 & 0.02$\pm$0.01   & \nodata         & \nodata         & \nodata     & 0.13$\pm$0.06 & \nodata      & \nodata     & \nodata & \nodata  & \nodata        & \nodata & \nodata     & \nodata     \\
          J=2--1 & \nodata         & 0.10$\pm$0.02   & \nodata         & 8.0$\pm$4.0 & \nodata       & \nodata      & \nodata     & \nodata & \nodata  & 0.4$\pm$0.2    & \nodata & 0.6$\pm$0.3 & \nodata     \\
          J=3--2 & \nodata         & \nodata         & 0.04$\pm$0.01   & \nodata     & \nodata       & \nodata      & \nodata     & \nodata & \nodata  & \nodata        & \nodata & \nodata     & \nodata     \\
C$^{18}$O J=2--1 & \nodata         & \nodata         & \nodata         & \nodata     & \nodata       & \nodata      & \nodata     & \nodata & \nodata  & \nodata        & \nodata & \nodata     & \nodata     \\
HCN J=1--0       & \nodata         & \nodata         & \nodata         & \nodata     & \nodata       & \nodata      & \nodata     & \nodata & \nodata  & \nodata        & \nodata & \nodata     & \nodata     \\
    J=3--2       & \nodata         & 0.3$\pm$0.1     & 0.2$\pm$0.1     & \nodata     & 3.0$\pm$2     & \nodata      & \nodata     & \nodata & \nodata  & \nodata        & \nodata & 1.9$\pm$0.5 & \nodata     \\ 
HCO$^+$ J=1--0   & \nodata         & \nodata         & \nodata         & \nodata     & \nodata       & \nodata      & \nodata     & \nodata & \nodata  & \nodata        & \nodata & \nodata     & \nodata     \\
        J=3--2   & \nodata         & 0.17$\pm$0.03   & 0.09$\pm$0.02   & \nodata     & 1.7$\pm$0.4   & \nodata      & \nodata     & \nodata & \nodata  & 0.5$\pm$0.3    & \nodata & \nodata     & \nodata     \\
        J=4--3   & \nodata         & \nodata         & \nodata         & \nodata     & \nodata       & \nodata      & \nodata     & \nodata & \nodata  & \nodata        & \nodata & \nodata     & \nodata     \\
\hline		   
\\[-0.15cm]

\multicolumn{14}{c}{Total}\\
$^{12}$CO J=1--0 & \nodata         & 0.3$\pm$0.2     & 0.3$\pm$0.2     & 40$\pm$20   & \nodata       & \nodata       & \nodata      & \nodata  & \nodata & \nodata      & \nodata & \nodata     & \nodata     \\
          J=2--1 & 2.0$\pm$1.0     & \nodata         & 0.3$\pm$0.2     & \nodata     & 8$\pm$3       & \nodata       & 94$\pm$6     & \nodata  & \nodata & 4.0$\pm$2.0  & \nodata & 9.0$\pm$4.0 & \nodata     \\
          J=3--2 & 4.0$\pm$2       & 4.0$\pm$2.0     & \nodata         & \nodata     & \nodata       & 41$\pm$2.0    & \nodata      & \nodata  & \nodata & 8.0$\pm$4.0  & \nodata & 16$\pm$1    & \nodata     \\
$^{13}$CO J=1--0 & 0.04$\pm$0.02   & \nodata         & \nodata         & \nodata     & 0.17$\pm$0.04 & 0.2$\pm$0.1   & \nodata      & \nodata  & \nodata & \nodata      & \nodata & \nodata     & \nodata     \\
          J=2--1 & \nodata         & 0.2$\pm$0.1     & \nodata         & 6.0$\pm$1.0 & \nodata       & 1.0$\pm$0.5   & \nodata      & \nodata  & \nodata & 0.3$\pm$0.2  & \nodata & 0.6$\pm$0.3 & \nodata     \\
          J=3--2 & \nodata         & \nodata         & 0.025$\pm$0.001 & 4.0$\pm$1.0 & 0.7$\pm$0.3   & \nodata       & \nodata      & \nodata  & \nodata & 0.2$\pm$0.1  & \nodata & 0.4$\pm$0.2 & \nodata     \\
C$^{18}$O J=2--1 & \nodata         & 0.011$\pm$0.001 & \nodata         & \nodata     & 0.2$\pm$0.1   & \nodata       & \nodata      & \nodata  & \nodata & \nodata      & \nodata & \nodata     & \nodata     \\
HCN J=1--0       & \nodata         & \nodata         & \nodata         & \nodata     & \nodata       & \nodata       & \nodata      & \nodata  & \nodata & \nodata      & \nodata & \nodata     & \nodata     \\
    J=2--1       & \nodata         & \nodata         & \nodata         & \nodata     & \nodata       & \nodata       & \nodata      & \nodata  & \nodata & \nodata      & \nodata & \nodata     & \nodata     \\
    J=3--2       & \nodata         & 0.3$\pm$0.1     & 0.12$\pm$0.01   & \nodata     & 3.0$\pm$0.3   & 5.0$\pm$3.0   & \nodata      & \nodata  & \nodata & \nodata      & \nodata & 1.9$\pm$0.2 & \nodata     \\
HCO$^+$ J=1--0   & \nodata         & \nodata         & \nodata         & \nodata     & \nodata       & \nodata       & \nodata      & \nodata  & \nodata & \nodata      & \nodata & \nodata     & \nodata     \\
        J=3--2   & \nodata         & 0.11$\pm$0.01   & 0.08$\pm$0.02   & \nodata     & 1.8$\pm$0.1   & 3.0$\pm$0.2   & \nodata      & \nodata  & \nodata & 0.5$\pm$0.3  & \nodata & \nodata     & 1.6$\pm$0.8 \\
        J=4--3   & \nodata         & \nodata         & \nodata         & \nodata     & \nodata       & 1.6$\pm$0.1   & \nodata      & \nodata  & \nodata & \nodata      & \nodata & 0.6$\pm$0.3 & \nodata     \\[-0.15cm]

\enddata 

\tablenotetext{$a$}{The line ratios were derived by spatially
averaging over the respective region from the velocity intergated line
ratio maps from Fig.~\ref{fig11.0}. The errors denote thereby the
standard deviation from the averaged values. Please note, that in some
cases, the line ratios might vary from those estimated from
Tables~\ref{tab4} and \ref{tab5}. See text for a discussion.}

\label{tab6}
\end{deluxetable}
\clearpage
\end{landscape}

\clearpage 

\clearpage 

\clearpage

\newpage

\clearpage 

\clearpage

\clearpage

\clearpage

\end{document}